\documentclass[a4paper,fleqn,usenatbib]{mn2e}
\pdfoutput=1
\usepackage{graphicx,amssymb, txfonts,pstricks}

\def\ha{H$\alpha$}

\def\oiii{[O\,{\sc iii}]}
\def\oi{[O\,{\sc i}]}
\def\sii{[S\,{\sc ii}]}
\def\nii{[N\,{\sc ii}]}
\def\kms{$\rm km\,s^{-1}$}
\def\hii{H\,{\sc ii}}
\def\arc{$^{\prime\prime}$}

\title[Dusty spirals vs. gas kinematics in LLAGN]{Dusty spirals versus gas kinematics in the inner kiloparsec of Four Low-Luminosity Active Galactic Nuclei}

\author[Brum et al.]{Carine Brum,$^{1}$\thanks{E-mail:
carinefisica@gmail.com} Rogemar A. Riffel$^{1}$, 
Thaisa Storchi-Bergmann$^{2}$, \newauthor Andrew Robinson$^{3}$, Allan Schnorr M{\"u}ller$^{2}$ and Davide Lena$^{4,5}$\\
$^{1}$ Universidade Federal de Santa Maria, Departamento de F\'\i sica, CCNE, 97105-900, Santa Maria, RS, Brazil\\
$^{2}$ Universidade Federal do Rio Grande do Sul, Instituto de F\'\i sica, CP 15051, Porto Alegre 91501-970, RS, Brazil\\
$^{3}$ Department of Physics, Rochester Institute of Technology, 84 Lomb Memorial Drive, Rochester, NY 14623, USA\\
$^{4}$ SRON, Netherlands Institute for Space Research, Sorbonnelaan 2, NL-3584 CA Utrecht, the Netherlands. \\
$^{5}$ Department of Astrophysics/IMAPP, Radboud University, Nijmegen, PO Box 9010, NL-6500 GL Nijmegen, the Netherlands.
}

\begin{document}


\pagerange{\pageref{firstpage}--\pageref{lastpage}} \pubyear{2016}

\maketitle

\label{firstpage}

\begin{abstract}

We used the Gemini Multi-Object Spectrograph (GMOS) Integral Field Unit (IFU) to
map the gas distribution, excitation and kinematics within the inner kiloparsec of four nearby
low-luminosity active galaxies: NGC3982, NGC4501, NGC2787 and NGC4450.
The observations cover the spectral range 5600--7000\,\AA\ at a velocity resolution of 120 \kms\
 and spatial resolution ranging from 50 to 70 pc at the galaxies. Extended emission in \ha,
[N\,{\sc ii}]$\lambda\lambda$6548,6583, [S\,{\sc ii}]\,$\lambda\lambda$6716,6730 over most of the field-of-view is observed for all galaxies,
while only NGC3982 shows [O\,{\sc i}]\,$\lambda$6300 extended emission. The \ha\ equivalent widths
($W_{H\alpha}$) combined with the [N\,{\sc ii}]/H$\alpha$  line ratios reveal that NGC3982 and NGC4450 harbor
Seyfert nuclei surrounded by regions with LINER excitation, while NGC2787
and NGC4501 harbor LINER nuclei. NGC3982 shows a partial ring of recent star-formation at 500 pc
from the nucleus, while in NGC4501 a region at 500pc west of the nucleus shows LINER excitation
but has been interpreted as an aging H{\sc ii}  region with the gas excitation dominated by
shocks from supernovae. The line-of-sight velocity field of the gas shows a rotation pattern for
all galaxies, with deviations from pure disk rotation observed in NGC3982, NGC 4501 and NGC 4450.
For NGC4501 and NGC4450, many of these deviations are spatially coincident with dust structures seen in optical continuum
images, leading to the interpretation that the deviations are due to shocks in the gas traced by the dust. A speculation is that these shocks lead to loss of angular momentum, allowing the gas to be transferred inwards to feed the AGN. In the case of NGC2787, instead of deviations in the rotation field, we see a misalignment of 40$^\circ$  between the orientation of the line of nodes of the gas rotation and the photometric major axis of the galaxy. Evidence of compact nuclear outflows are seen in NGC4501 and NGC4450.

\end{abstract}


\section{Introduction}

Understanding how mass is transferred from galactic scales down to nuclear scales to feed the
super-massive black hole (SMBH) in the nuclei of galaxies has been the goal of many theoretical studies and simulations 
\citep{shlosman, maciejewski, emsellem, knapen, emsellem2006,emsellem15,li15,capelo17}. These works have shown that non-axisymmetric potentials efficiently promote gas inflow toward the inner kiloparsec of galaxies \citep[e.g.][]{engl2004}, resulting in a gas reservoir that can trigger and maintain an Active Galactic Nucleus (AGN) and/or nuclear star formation.

Nuclear bars and associated spiral arms are indeed frequently observed in the inner kiloparsec of active galaxies
\citep[e.g.][]{erwin, pogge, laine}. \citet{slopes} found a strong correlation between the presence of the nuclear dusty structures (filaments, spirals, discs and bars) and  nuclear activity in a sample of early-type galaxies, suggesting that a reservoir of gas and dust is a necessary condition for a galaxy to harbor an AGN. This correlation between the presence of dusty structures and nuclear activity supports the hypothesis that these structures represent a fueling mechanism for the SMBH, allowing the gas to loose angular momentum and stream towards the center of the galaxies.

Previous studies by our group \citep[e.g.][]{fathi06, thaisa07, allan11, allan14, allan14b}, have revealed kinematic 
features associated with nuclear spirals, bars or filaments, that are consistent with gas inflow
 to the inner tens of parsecs of active galaxies. Motivated by these results, we have mapped the gaseous kinematics of four nearby AGNs showing dusty nuclear spirals, 
with the goal of looking for correlations between these spirals and the gas kinematics. 
 The galaxies NGC\,3982, NGC\,4501, NGC\,2787 and NGC\,4450 were selected from the work by \citet{slopes}, that was based mostly on low-luminosity AGNs. This study is part of a larger project in which we are obtaining optical integral field spectroscopic observations of a complete X-ray selected sample with the aim of investigating feeding and feedback mechanisms over a range in AGN luminosity \citep[e.g.][]{allan14,lena15}.

This work is organized as follows: Section 2 presents a description of the observations and data reduction procedure, Sec. 3 shows the emission-line flux distributions, line-ratio maps, velocity fields and velocity dispersion maps. In Sec. 4 we model the velocity fields and in Sec. 5 we discuss the results for each galaxy. Finally, in Sec. 6, we present the main conclusions of this work.

\section{Observations and Data Reduction}\label{obs}

As pointed out above, the four active galaxies of this study were selected from the sample of \citet{slopes} by showing dusty nuclear spirals in Hubble Space Telescope (HST) optical images through the filter F606W, and revealed in  ``structure maps'' that are aimed to enhance fine structural features in single-filter images \citep{pogge}.  
The observations were obtained using the  Gemini Multi-Object Spectrograph Integral Field Unit \citep[GMOS-IFU,][]{allington-smith02,hook04} at the Gemini North Telescope in 2007, 2008 and 2011.

We have observed the wavelength range 5600-7000\,\AA, which includes the strongest emission lines, as  
H$\alpha$, [N\,{\sc ii}]\,$\lambda\lambda\,6548,6583$,  [S\,{\sc ii}]\,$\lambda\lambda\,6716,6730$ 
and [O\,{\sc i}]\,$\lambda$6300, using the IFU in the two slit mode. The R400 grating was used in combination with the {\it r}(530\,nm) filter, resulting in a spectral resolution of 2.5-2.7~\AA, as obtained from the full-width at half maximum (FWHM) of arc lamp lines used to wavelength calibrate the spectra, translating to $\sim$100-125\,km\,s$^{-1}$ in velocity space.

The data comprise three adjacent IFU fields for NGC\,3982, NGC\,4501 and NGC\,4450, each one covering 5$^{\prime\prime}$ $\times$ 7$^{\prime\prime}$, while for NGC\,2787 we used two adjacent IFU fields. In order to remove cosmic rays and bad pixels small spatial offsets were performed between individual exposures at each position. The final Field of View (FoV), obtained after mosaicking the individual cubes for each galaxy are approximately $7^{\prime\prime}\times15^{\prime\prime}$  for NGC\,3982 and NGC\,4501, $7^{\prime\prime}\times9^{\prime\prime}$ for NGC\,2787 and $20^{\prime\prime}\times5^{\prime\prime}$ for NGC\,4450, with the longest dimension of the FoV oriented along the major axis of each galaxy.
The total exposure time for each galaxy ranges from 75 to 82 minutes.
Table~\ref{ob} shows the log of the observations, as well as relevant information on each galaxy. 

\begin{table*}
\scriptsize
\caption{Observing log and properties of the sample}
\begin{tabular}{c c c c c c c c c c c}
\hline
\hline
Galaxy 	& Nuc. act. & Hubble Type &  d\,(Mpc) & Obs. Date  &  Exp. Time$^1$  & NFoV	& FoV   &      Spec. Res. & Spat. Res. &    Gemini Project\\
\hline
NGC\,3982 & Seyfert 2 & SAB(r)b$^{2}$ & 17.0$^{2}$ & 01/01/2007 & 3$\times$520& 3 & 7\farcs1$\times$15\farcs3 & 2.7\AA & 70\,pc (0\farcs85) & GN-2006B-Q-94 \\ 
NGC\,4501 & Seyfert 2 & SA(rs)b$^{3}$ & 16.8$^{2}$ & 02/14/2008 & 3$\times$500& 3 &7\farcs3$\times$15\farcs5 & 2.5$\AA$ & 60\,pc (0\farcs75) & GN-2008A-Q-8 \\ 
NGC\,2787 & LINER & SB(r)0$^{+2}$ & 13.0$^{2}$ & 01/27/2011 & 4$\times$615 &2& 7\farcs0$\times$9\farcs2 & 2.5$\AA$ & 50\,pc (0\farcs80)  & GN-2011A-Q-85 \\
NGC\,4450 & LINER & SAab$^{2}$ & 16.8$^{2}$ & 01/03/2007 & 3$\times$520 &3& 20\farcs1$\times$5\farcs5 & 2.7$\AA$ & 65\,pc (0\farcs80)  & GN-2006B-Q-94 \\ 
\hline
\multicolumn{11}{l}{(1) Exposure time per FoV; (2) \citet{ho97}. (3) \citet{devaucouleurs91}.}
\end{tabular}
\label{ob}
\end{table*}

Data reduction was performed using the IRAF\footnote{IRAF is distributed by National Optical Astronomy Observatories, which are operated by the Association of Universities for Research in Astronomy, Inc., under cooperative agreement with the National Science Foundation} packages provided by the Gemini 
Observatory and specifically developed for the GMOS instrument. 
First, the bias was subtracted from each image, followed by flat-fielding and trimming of the spectra. The wavelength calibration was applied to the science data using as reference the spectra of Arc lamps, followed by subtraction of the underlying sky emission. To obtain a relative flux calibration we constructed a sensitivity function using the spectra of standard stars provided by the observatory as default calibrations. Feige\,66 was used as standard star for NGC\,4501, while Feige 34 was used for NGC\,2787 and BD+28d4211 was used for NGC\,3982 and NGC\,4450. 
Finally, separate data cubes were obtained for each exposure, which were then aligned and median combined to a single cube for each galaxy. 
All individual cubes were created with 0\farcs05 square spaxels during the data reduction, but for NGC\,4450 the final
data cube was rebinned to 0\farcs15 square spaxels in order to increase the signal to noise ratio of the spectra and allow the fitting of the emission-lines profiles at locations away from the nucleus.


The spatial resolution of the final  data cubes, presented in Table~\ref{ob}, are in  the range 50--70 pc and were obtained from average FWHM of the flux distribution of field stars seen in the acquisition image at the $r$ band, and adopting the distances quoted in the forth column of the same table. The uncertainty in the spatial resolution is about 10\,\% for all galaxies, estimated as the standard deviation of the average FWHM.

\section{Results}
 
\begin{figure*}
\includegraphics[scale=0.25]{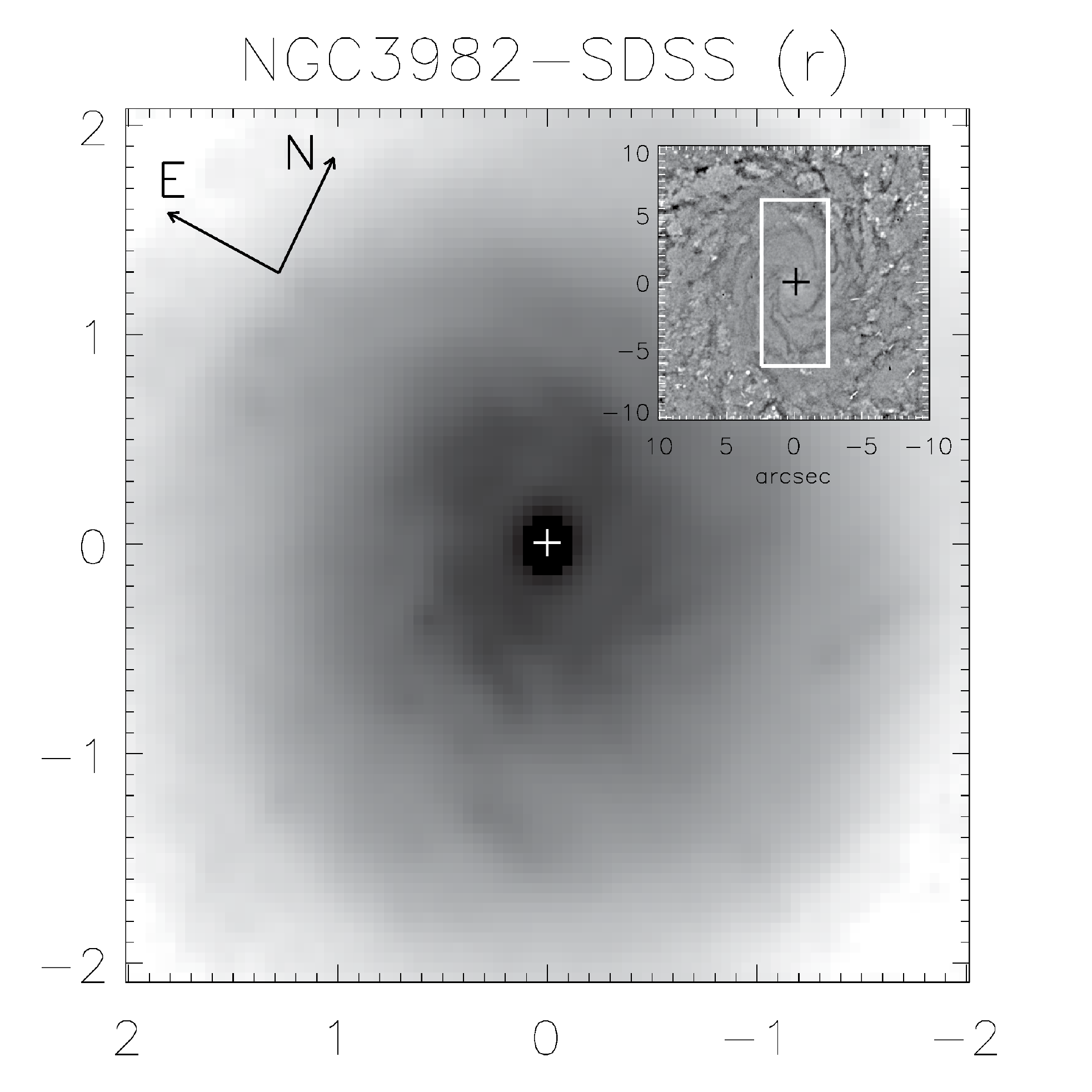}
\includegraphics[scale=0.34]{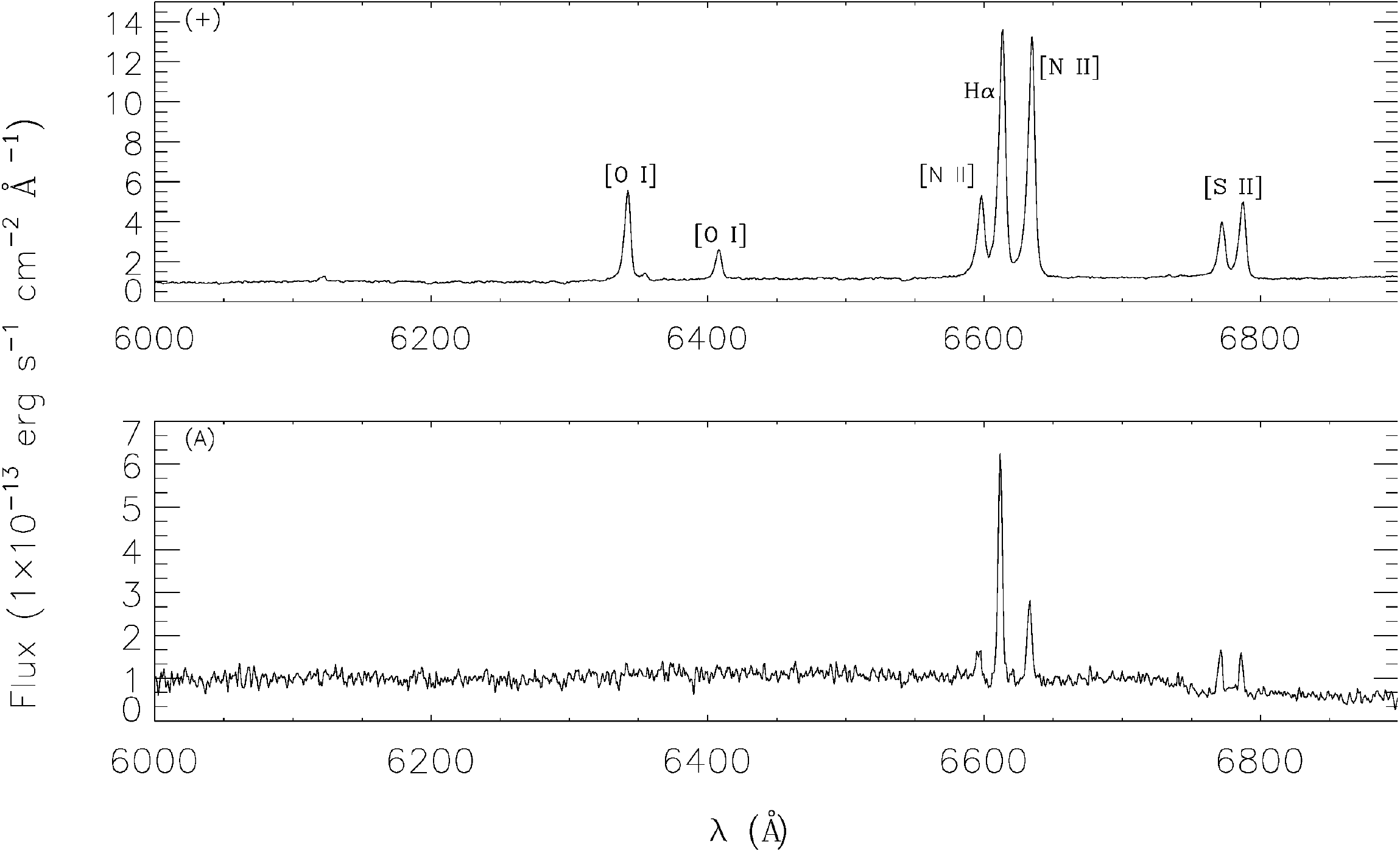}
\includegraphics[scale=0.25]{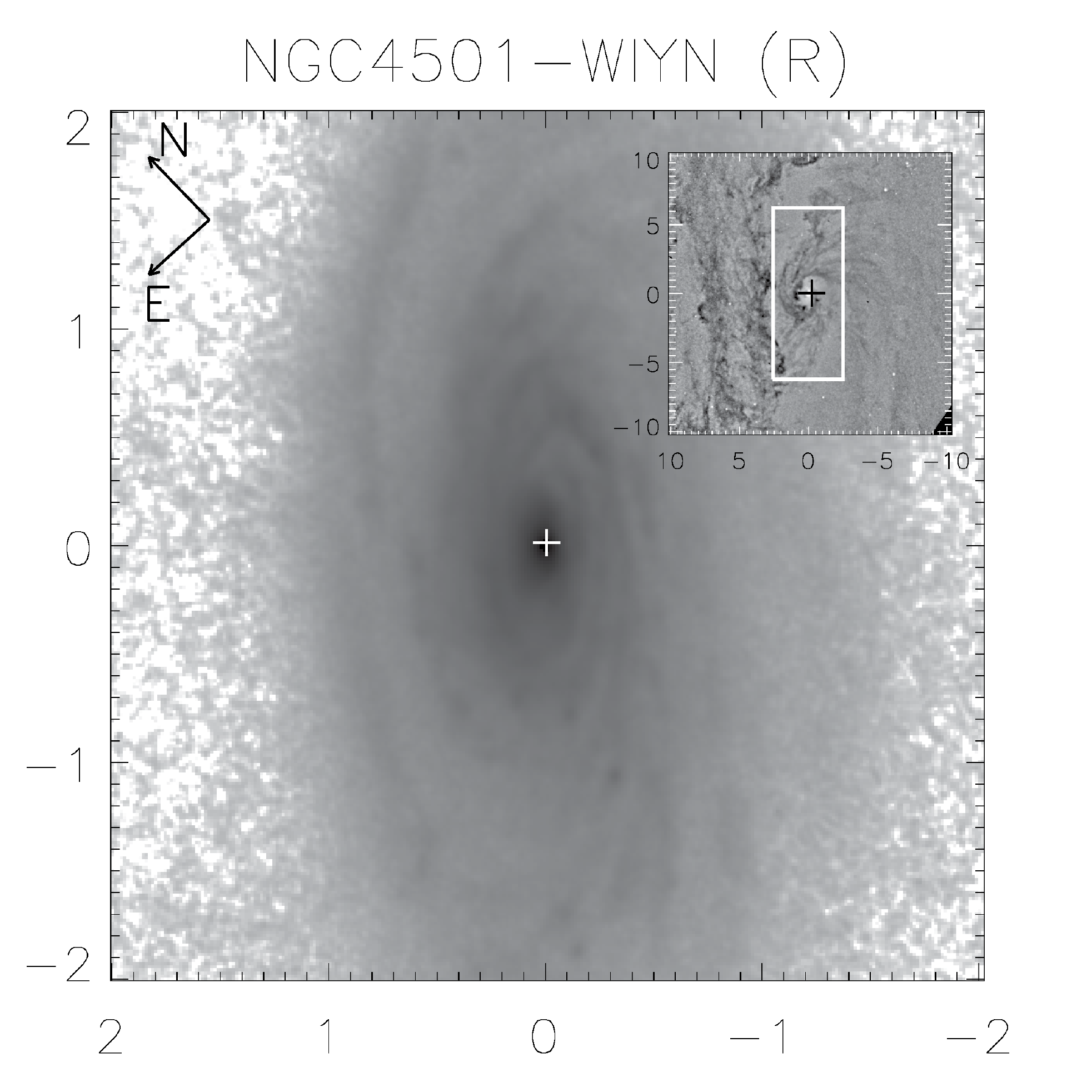}
\includegraphics[scale=0.34]{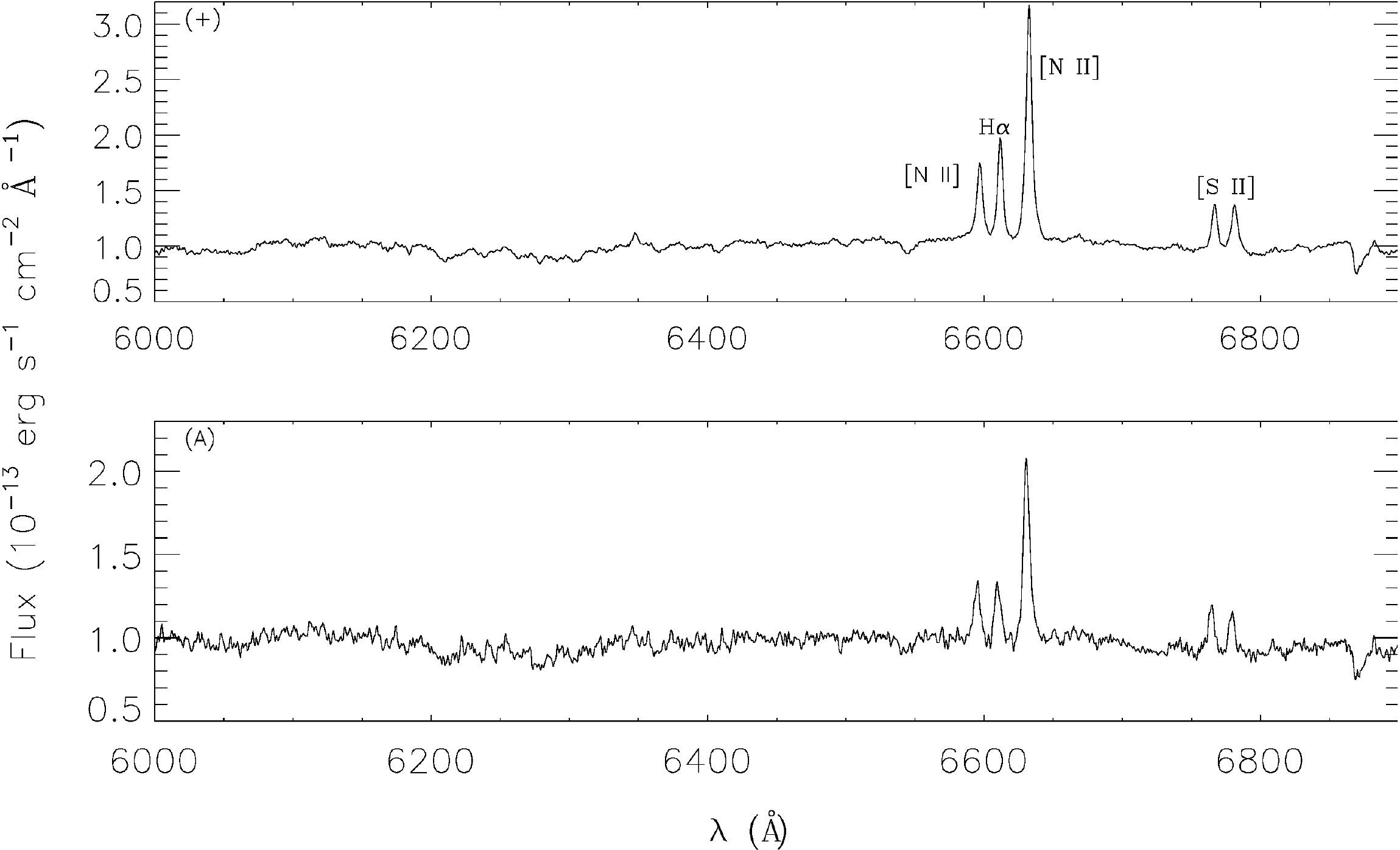}
\includegraphics[scale=0.25]{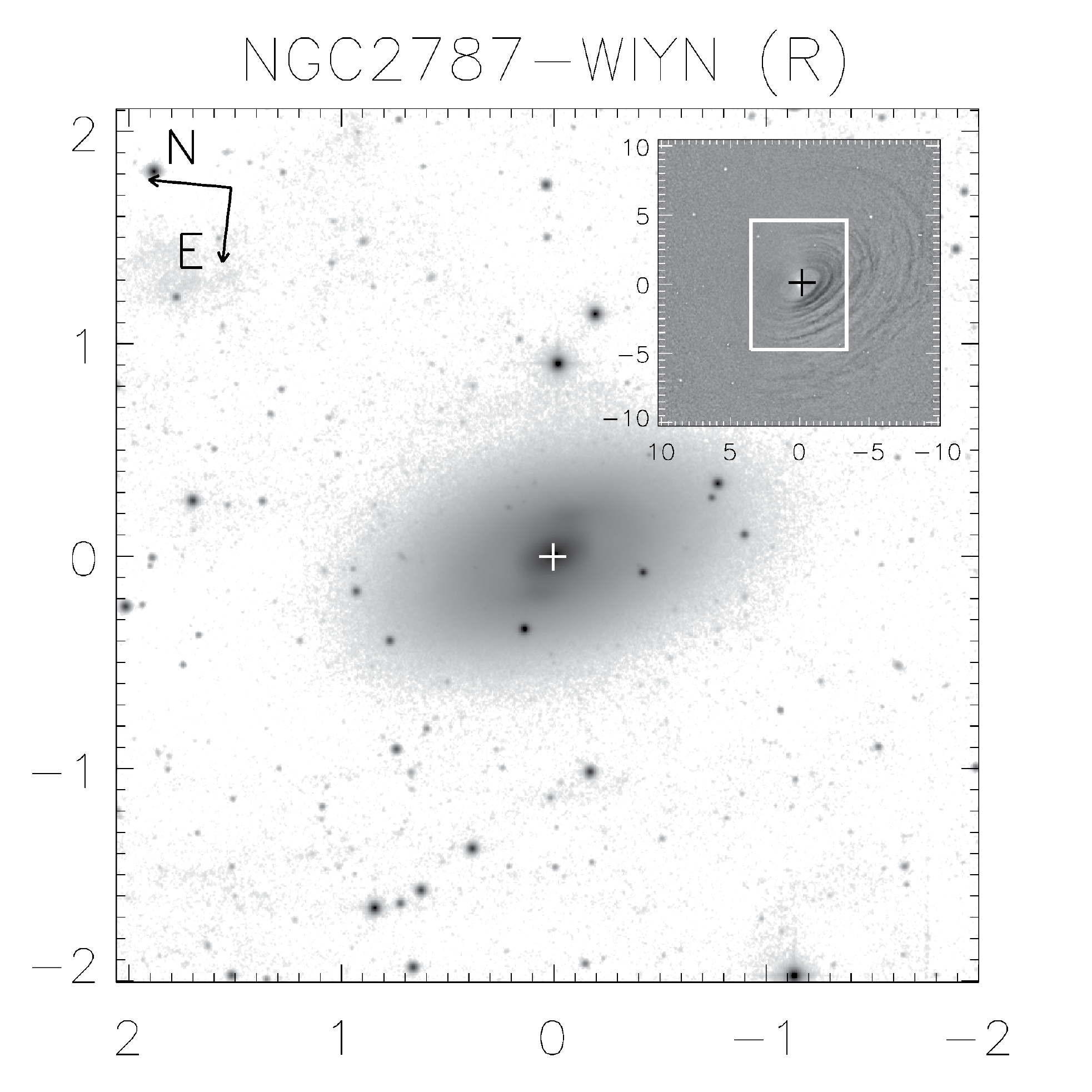}
\includegraphics[scale=0.34]{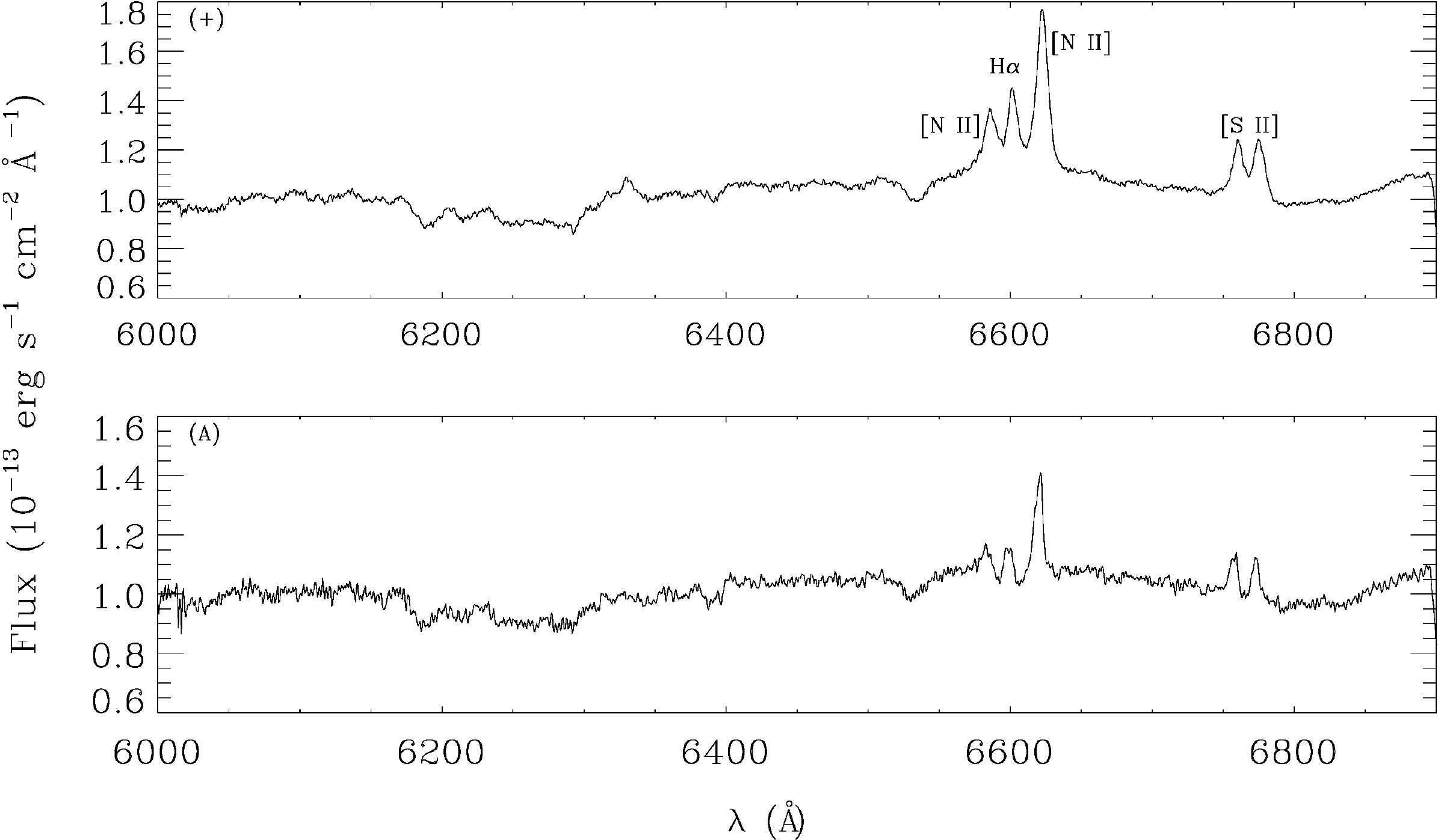} 
\includegraphics[scale=0.25]{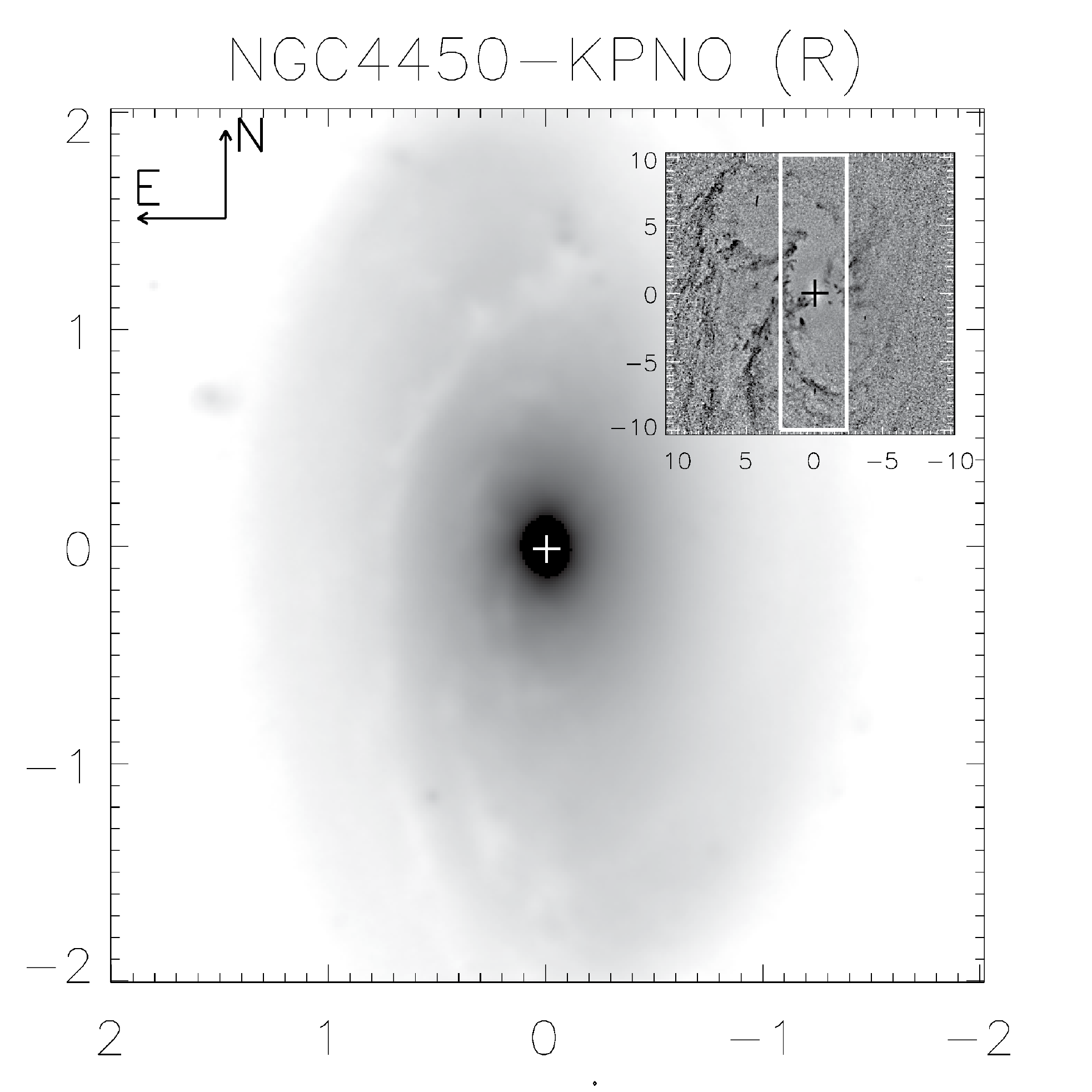}
\includegraphics[scale=0.34]{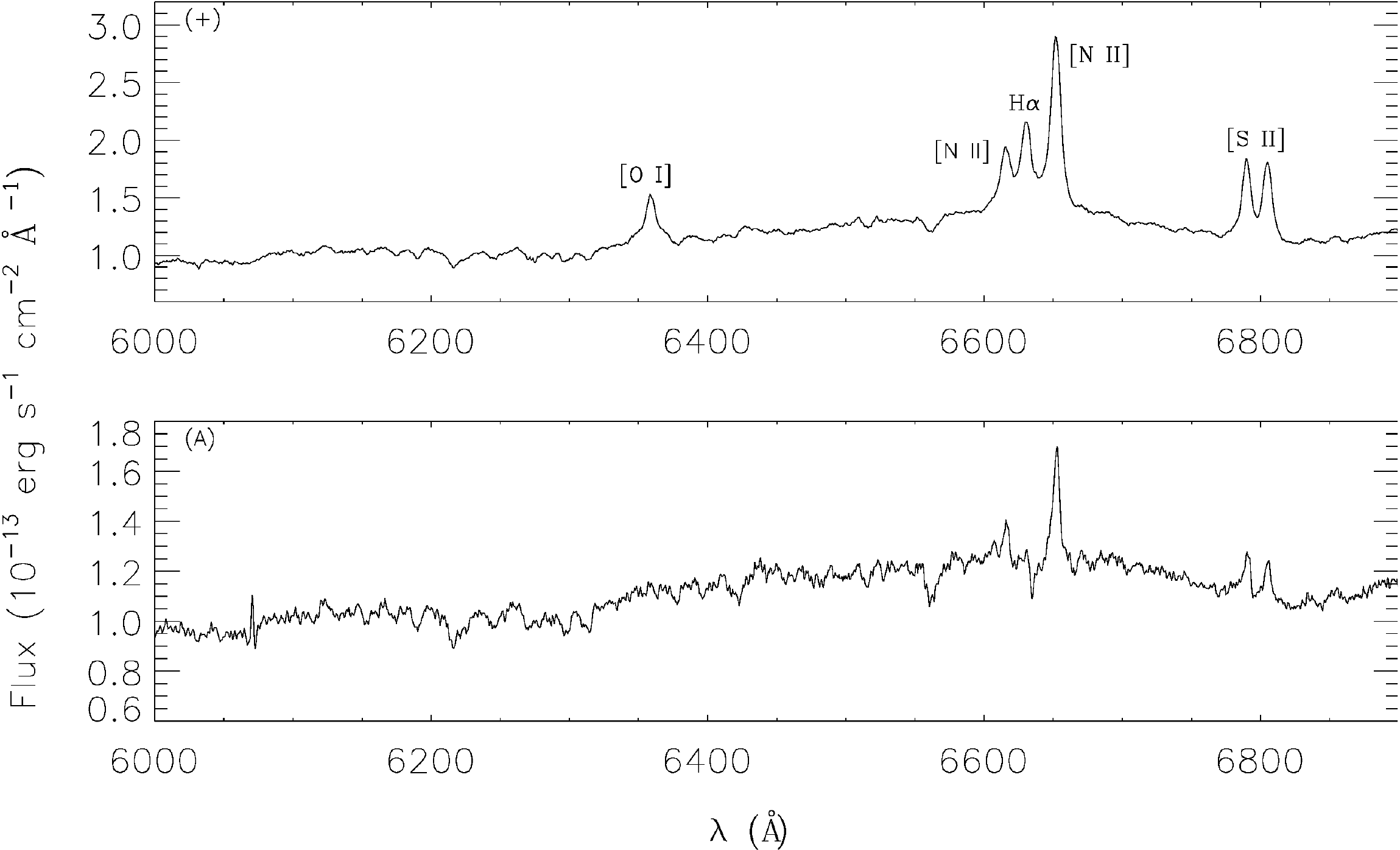} 
 \caption{Left panels: large-scale image obtained from SDSS \citep{baillard11} for NGC\,3982, from \citet{koopmann01} for NGC\,4501 and NGC\,2787 and from \citet{young96} for NGC\,4450. Insert in the top right corner of in each left panel: structure map obtained from an HST-WFPC2 F606W image \citep{slopes} of the inner 20$^{\prime\prime} \times 20^{\prime\prime}$. 
Right panels show spectra from the nucleus (labeled as `+'), and positions ``A'' identified at the flux distribution image of in Figs.~\ref{all-figs4501} and \ref{figs45}.  These spectra were extracted within a circular aperture of 0\farcs25 radius.}
\label{maps}
\end{figure*}

Figure~\ref{maps} shows a large scale image and typical spectra for each galaxy. At the top-right corner of each image, we show as an insert the structure map of the inner 20$^{\prime\prime}$$\times$20$^{\prime\prime}$. 
The orientation is shown by the arrows in the top-left corner of each image. The central box at the HST structure map shows the IFU FoV. These structure maps were constructed following \citet{slopes} using HST broad band images obtained through the F606W filter \citep[from ][]{malkan98} and are aimed to highlight dust structures present in the inner region of the galaxies. 
The structure maps reveal dust structures for all galaxies: NGC\,3892 clearly shows dust spiral structures from 10$^{\prime\prime}$ down to the nucleus; NGC\,4501 shows much more dust to the northeast than to the southwest; NGC\,2787 shows elliptical dusty partial rings that seem to be concentric and NGC\,4450 presents a complex dust distribution with more dust to the east of the nucleus, and a dusty blob to the northwest.

The right panels of Fig~\ref{maps} show two spectra for each galaxy obtained within a circular aperture of 0\farcs25 radius (corresponding to 5 pixels) for NGC\,3982, NGC\,4501 and NGC\,2787,
and 0\farcs45 radius (corresponding to 3 pixels) for NGC\,4450. The first spectrum correspond to the nucleus and the other were obtained for the position labeled as ``A" in the flux distributions maps of Figures~\ref{all-figs3982}--\ref{all-figs4450}, chosen to represent typical extra-nuclear spectra. The strongest emission lines are identified.

\subsection{Measurements}

In order to measure the emission-line flux distributions and gas kinematics, we fitted the line profiles of H$\alpha$+[N\,{\sc ii}]\,$\lambda\lambda$\,6548,6584,  [S\,{\sc ii}]\,$\lambda\lambda$\,6717,6731 and [O\,{\sc i}]$\lambda6300$ by single Gaussian curves using a modified version of the {\sc profit} routine \citep{profit}. This routine performs the fit of the observed profile using the MPFITFUN routine \citep{mark09}, via a non-linear least-squares fit.  In order to reduce the number of free parameters, we adopted the following constraints: the [N\,{\sc ii}]$+$H$\alpha$ emission lines were fitted by keeping tied the kinematics of the [N\,{\sc ii}] lines and fixing the [N\,{\sc ii}]$\lambda6563$/[N\,{\sc ii}]$\lambda6548$ intensity ratio to its theoretical value (3). The [S\,{\sc ii}] doublet was fitted by keeping the kinematics of the two lines tied, while the [O\,{\sc i}]\,$\lambda6300$ was fitted individually with all parameters free.  For the H$\alpha$ line all parameters were allowed to vary independently. In all cases, the continuum emission was fitted by a linear equation, as the spectral range of each line fit was small. 

NGC\,4450 presents a known broad double-peaked H$\alpha$ line \citep{ho00}, seen also in the nuclear spectrum of our GMOS data (Fig.~\ref{maps}). To take this broad double-peaked emission into account during the fit, we have included two additional broad components at locations closer to the nucleus than 1$^{\prime\prime}$. As the Broad Line Region is not resolved, the width and central wavelength of each broad component were kept fixed to the values obtained by fitting the nuclear spectrum shown in Fig.~\ref{maps}, while the amplitude of each Gaussian was allowed to vary during the fit. 

\begin{figure*}
 \centering
  \includegraphics[scale=0.65]{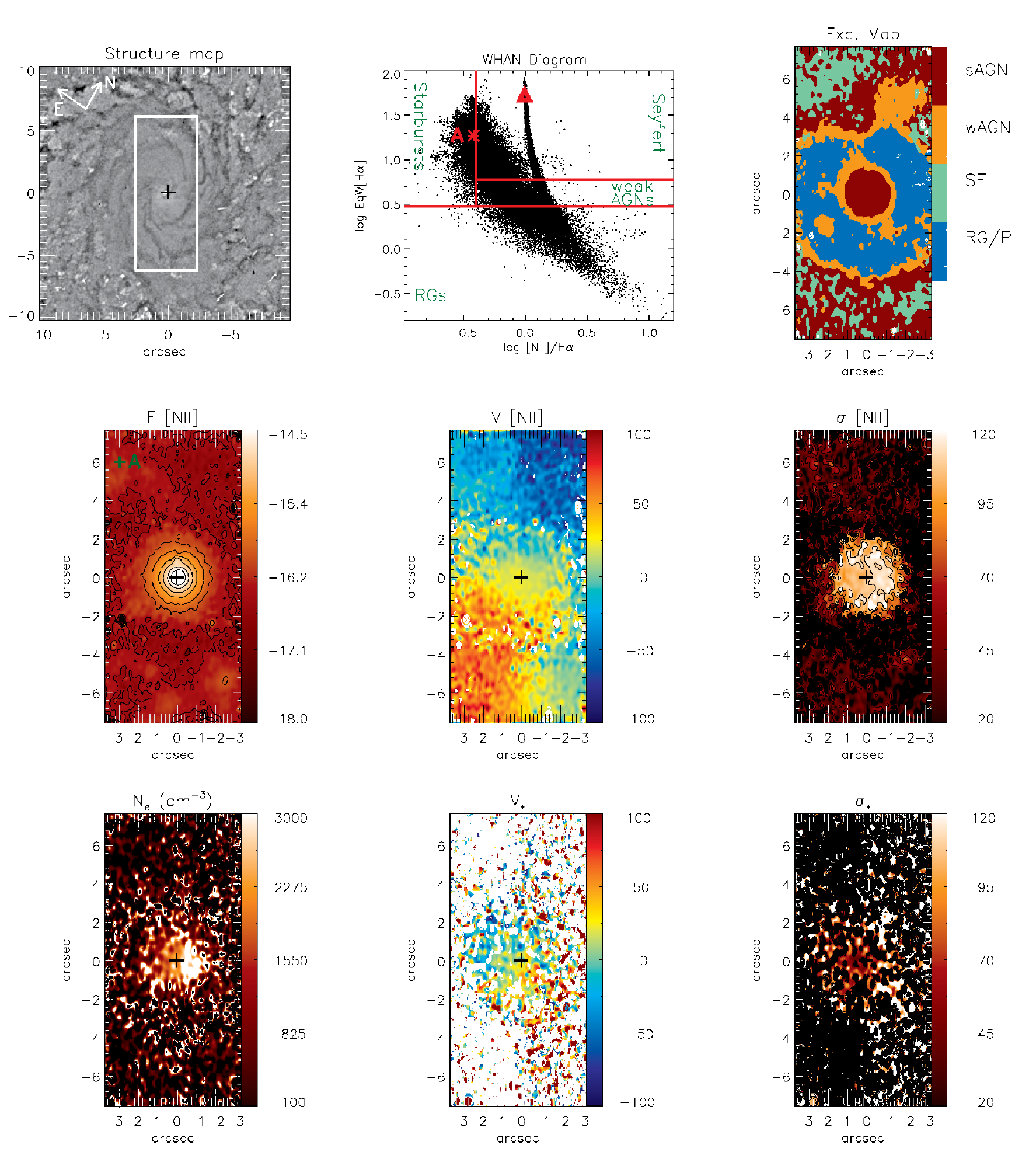} 
   \caption{NGC\,3982 - Top panels: structure map of the HST F606W image of the inner 20$^{\prime\prime}\times$20$^{\prime\prime}$; the WHAN diagram showing the region occupied by the distinct excitation classes, where  the \ha\ equivalent width is shown in \AA; the red triangle corresponds to the nucleus and the asterisks represents position A (see spectra in Fig.~\ref{maps}). Middle panels: flux distribution in logarithmic units of  erg\,s$^{-1}$\,cm$^{-2}$; line-of-sight velocity field; and velocity dispersion map for the [N\,{\sc ii}]\,$\lambda$6583 emission line, both in units of \kms. Bottom panels: gas density (cm$^{-3}$) derived from the [S\,{\sc ii}] line ratio, stellar centroid velocity field (km\,s$^{-1}$) and stellar velocity dispersion (km\,s$^{-1}$). The central crosses mark the location of the nucleus.}
\label{all-figs3982}
\end{figure*}

\begin{figure*}
 \centering
  \includegraphics[scale=0.65]{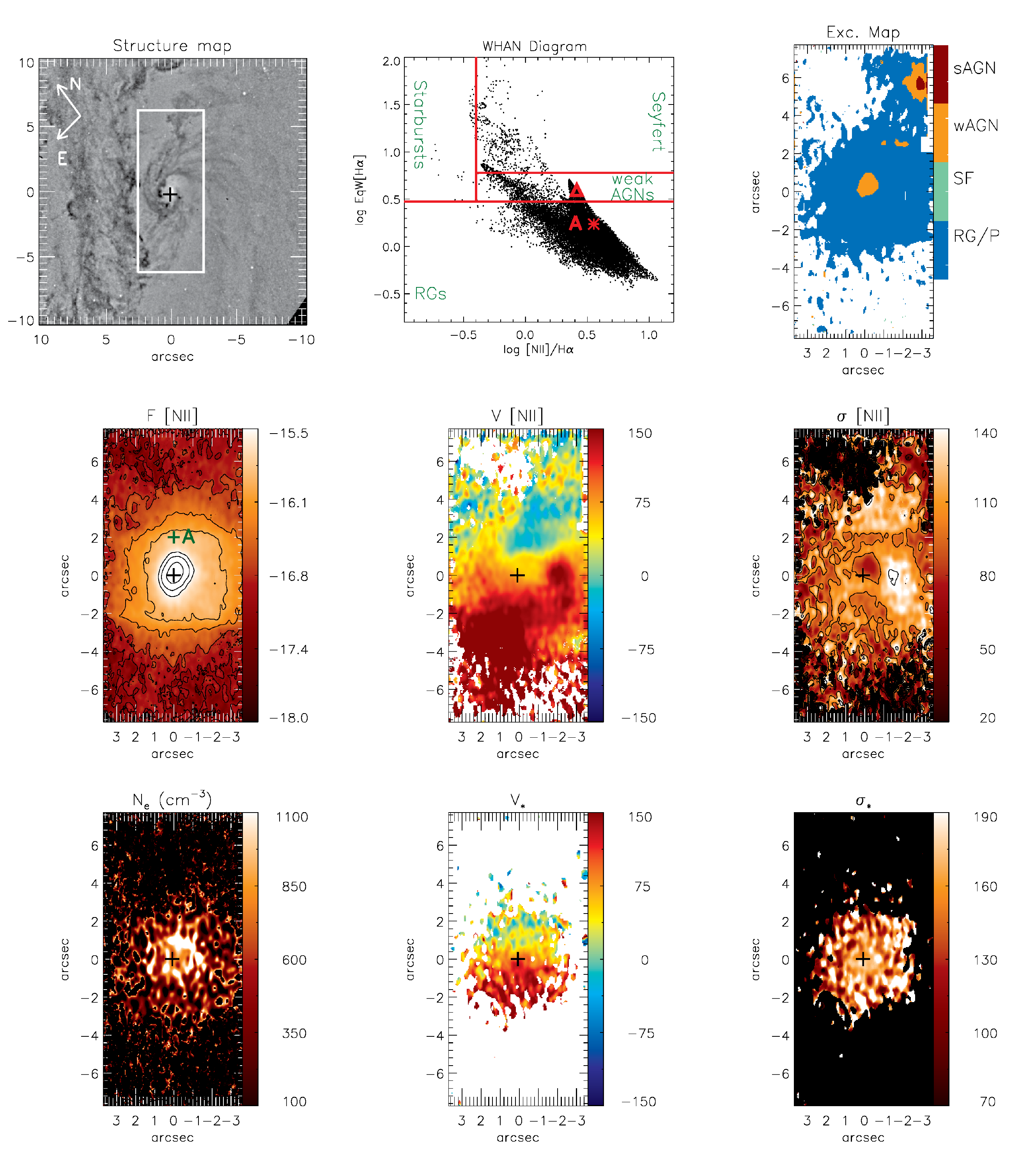} 
   \caption{Same as Fig. \ref{all-figs3982} for NGC\,4501.} 
\label{all-figs4501}
\end{figure*}

\begin{figure*}
 \centering
  \includegraphics[scale=0.6]{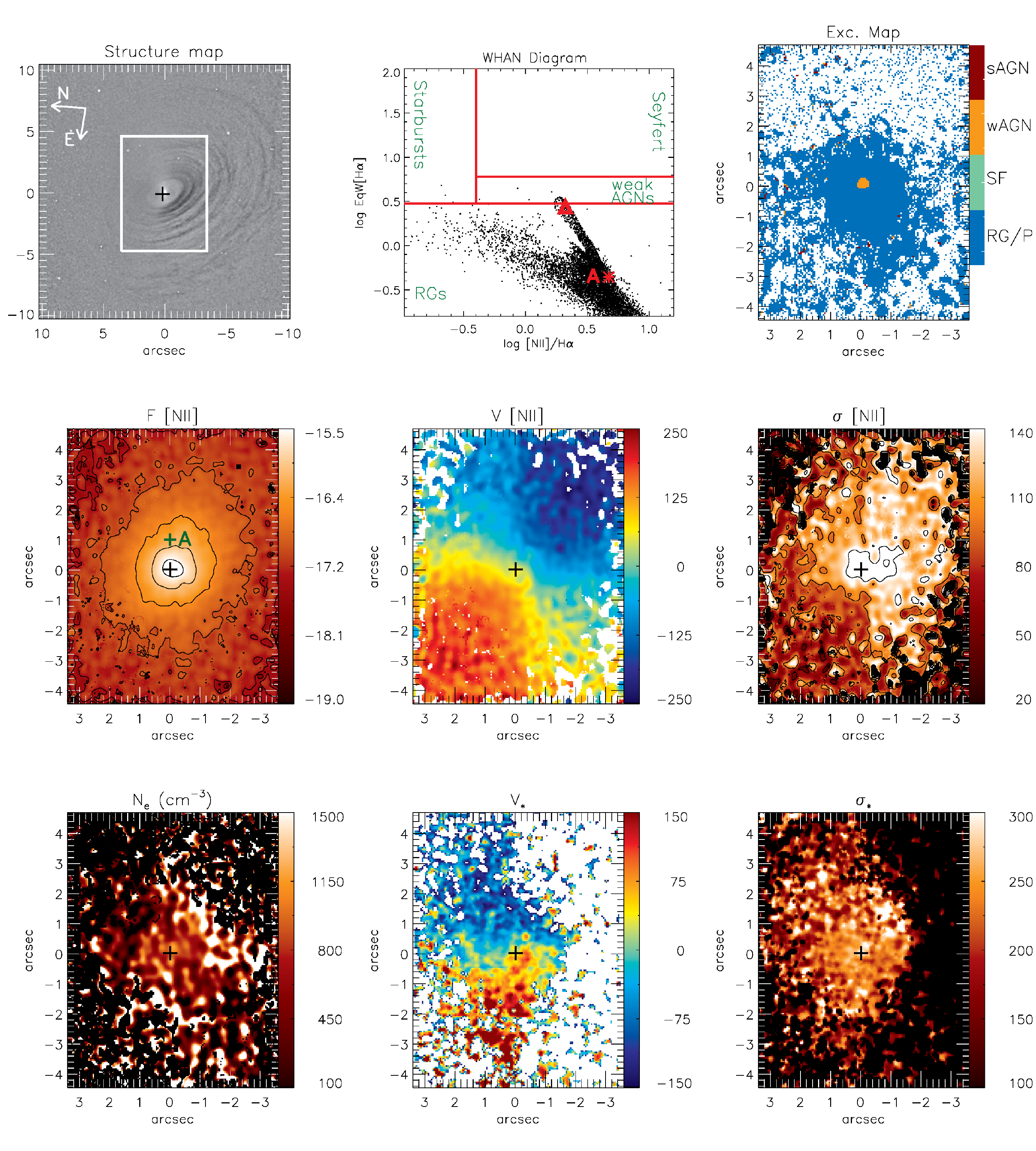} 
   \caption{Same as Fig.~\ref{all-figs3982} for NGC\,2787.}
\label{all-figs2787}
\end{figure*}

\begin{figure*}
 \centering
  \includegraphics[scale=0.7]{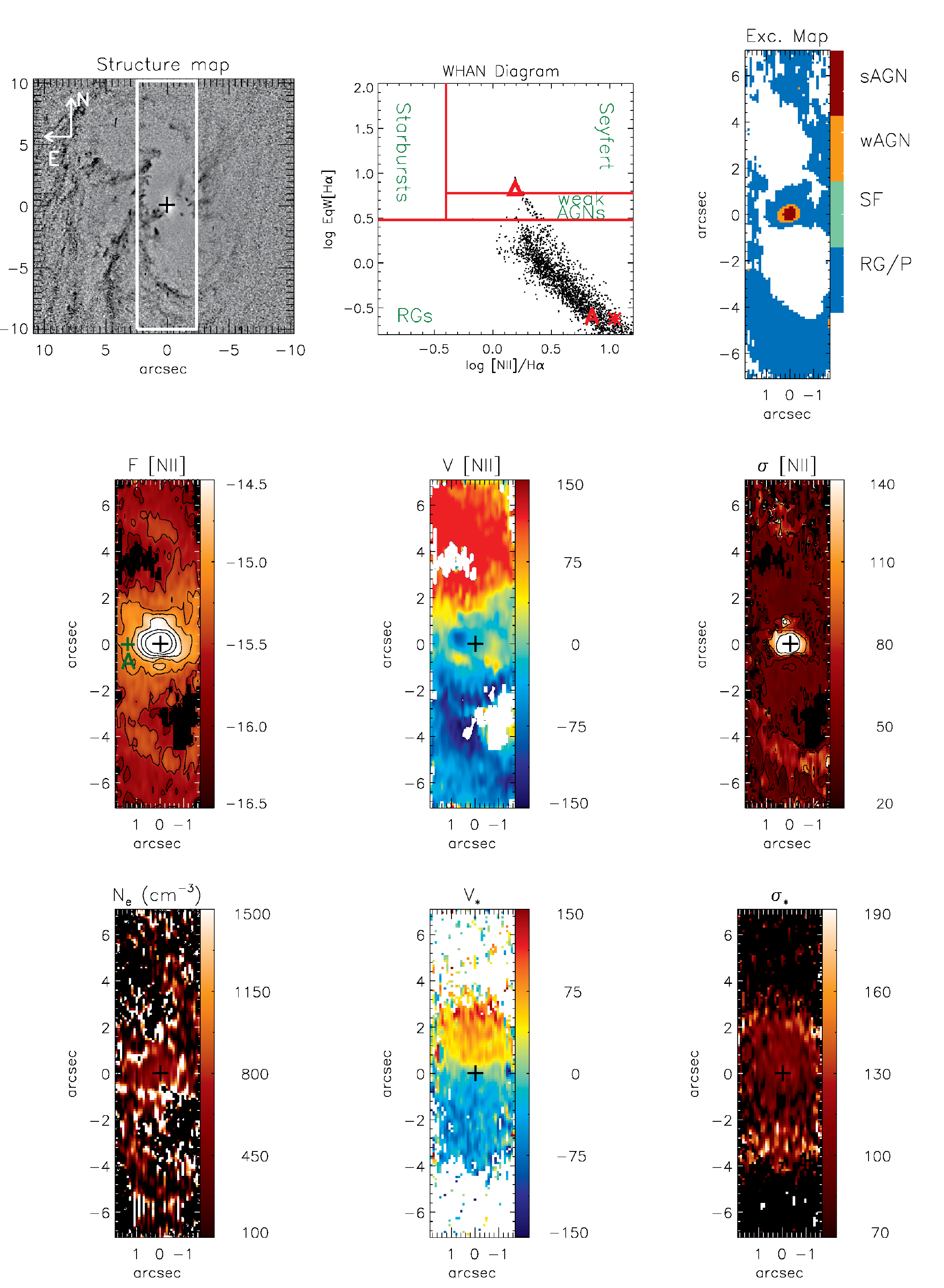} 
   \caption{Same as Fig. \ref{all-figs3982} for NGC\,4450.} 
\label{all-figs4450}
\end{figure*}

The fitting routine outputs a data cube with the emission-line fluxes, gas velocity, and velocity dispersion, as well as their corresponding uncertainties and $\chi^2$ map. These cubes were used to construct two dimensional maps for these physical properties, presented in Figures~\ref{all-figs3982} - \ref{all-figs4450}, for NGC\,3982, NGC\,4501, NGC\,2787 and NGC\,4450, respectively, together with the HST structure map, an excitation diagram, excitation map and electronic density.



\subsection{WHAN diagram and excitation map}

In order to map the gas excitation, line-ratio diagnostic diagrams are frequently used, the most popular of them being
the BPT diagrams \citep{bpt}. Integral field spectroscopy allows the construction of two-dimensional diagnostic diagrams \citep{sarzi10,sanchez15,colina15,belfiore16}. As our observations do not cover the \oiii$\lambda5007$ and H$\beta$ lines, we use an alternative diagnostic diagram recently proposed by \citet{cid10}, that
makes use only of the \ha\ and \nii$\lambda65863$ emission lines. This diagram plots the \ha\ equivalent width against the [N\,{\sc ii}]/H$\alpha$ line ratio, and is usually referred to as the WHAN diagram \citep{cid10}. 
The WHAN diagram allows the separation of Starbursts, Seyfert galaxies (or strong AGN: sAGN), weak AGN (wAGN, defined as having Ha equivalent widths W (H$\alpha$) between 3\AA\ and 6\AA), and retired galaxies (RGs), that is galaxies having W(H$\alpha$) smaller than 3\AA, not active, but displaying weak emission lines produced by radiation from post-AGB stars. A particular advantage of the WHAN diagram is to allow the separation between wAGN and RG, that overlap in the LINER region of traditional BPT diagnostic diagrams.

The spatially resolved WHAN diagnostic diagrams for each galaxy are shown in the top-central panels of Figures~\ref{all-figs3982}--\ref{all-figs4450} and the top-right panels show the resulting excitation maps: distinct regions of the FoV color-coded according to the excitation derived from the WHAN diagram. White regions in the excitation maps correspond to locations where we could not fit one or both emission lines. The color bar shows the identification of the distinct excitation classes. 

The excitation map for NGC\,3982 shows sAGN values within the inner 1$^{\prime\prime}$ (82 pc), while SF excitation regime is observed in a ring at 4-6$^{\prime\prime}$ from the nucleus and a mixture of RG, sAGN and wAGN is observed in between these regions. The excitation map of NGC\,4501 and NGC\,2787 show  unresolved regions of wAGN excitation at their nuclei, surrounded by RG excitation regions. A similar behavior is observed for NGC\,4450, but showing sAGN values at the nucleus. In addition, NGC\,4501 presents sAGN values within an unresolved region at $\sim$6\arc\ west of the nucleus (close to the border of the FoV).

\subsection{Flux distributions}

The flux distributions in the  \nii$\lambda6583$ emission line for each galaxy is shown in the middle-left panels of Figures~\ref{all-figs3982}--\ref{all-figs4450}. Black regions represent masked locations where the uncertainty in the flux is larger than 30\%,  and we were not able to get good fits of the line profiles due to the low S/N ratio or non-detection of the corresponding emission line.  
Similar maps for the \ha, \sii\ and \oi\ emission lines are shown in appendix A (Figures  \ref{figs39}--\ref{figs44}). The [O\,{\sc i}]\,$\lambda$6300\,\AA\ flux maps show extended emission only for NGC\,3982 and we thus do not show the corresponding flux maps for the other galaxies.

All galaxies present the gas emission peak at the nucleus for all emission lines.

The NGC\,3982  flux maps for \ha\ and \nii\ show extended emission over the whole GMOS-IFU FoV (up to 574\,pc (7$^{\prime\prime}$) from the nucleus), with a partial ring of enhanced gas emission surrounding the nucleus at 328 - 492\,pc ($4-6^{\prime\prime}$) from it, attributed to star forming regions. The \sii\ flux distribution also shows emission associated to the ring, but most of the \sii\ emission is concentrated within $r=164-328\,$pc (2-4$^{\prime\prime}$) from the nucleus. The \oi\ emission is observed only within $r=123\,$pc (1.5$^{\prime\prime}$) from the nucleus.


NGC\,4501 presents extended emission for \ha\, and \nii\  up to 486\,pc (6$^{\prime\prime}$) from the nucleus. The \sii\ flux map is more concentrated, with emission seen only within the inner 162 -- 243\,pc (2-3$^{\prime\prime}$).
All maps show the most extended emission along the northwest-southeast direction, approximately along the the major axis of the galaxy as seen in Fig.~\ref{maps}. In addition, the \ha\ map presents a small region with
enhanced emission at $\sim$486\,pc ($6^{\prime\prime}$) south-west of the nucleus attributed to an \hii\ region.

Extended \nii\ and \sii\ emission over the whole FoV -- up to 315\,pc (5$^{\prime\prime}$) from the nucleus is observed for NGC\,2787, while the \ha\ emission is more concentrated within the inner 126 -- 189\,pc ($2-3$\arc).

For NGC\,4450, the highest intensity levels show flux distributions slightly more elongated along the east-west direction, approximately perpendicular to the orientation of the major axis of the galaxy. At locations beyond 162 -- 243\,pc (2-3$^{\prime\prime}$) from the nucleus, the \nii\ and \ha\ emission show two spiral arms, one to the north and another to the south of the nucleus, extending to up to 810\,pc (10\arc) from it.

\subsection{Velocity fields}





The \nii\ velocity field for each galaxy is presented in the mid-central panels of  Figures~\ref{all-figs3982}--\ref{all-figs4450}, with white regions corresponding to masked locations due to bad fits. Similar maps for the \ha, \sii\ and \oi\ emission lines are shown in Figures  \ref{figs39}--\ref{figs44}.

The \ha\ and \nii\ velocity fields for NGC\,3982  present a rotation pattern with blueshifts observed to the north (thus this side is approaching) and redshifts to the south (and this side is receding), with a projected 
velocity amplitude of $\approx$ 100 km\,s$^{-1}$. 


NGC\,4501 also presents velocity fields consistent with gas rotating in the plane of the galaxy with blueshifts to the west and redshifts to the east, also showing deviations from pure rotation indicating the presence of non-circular motions at some locations. The velocity fields for all lines are similar, with deviations from rotation including excess redshifts at $\sim$162\,pc (2\arc) southwest of the nucleus, along the minor axis of the galaxy (where the velocities reach values of up to 150\,\kms) and  excess redshifts in a region marginally resolved at $\sim$486\,pc (6\arc) west of the nucleus. The origin of these structures will be further discussed in Sec.~\ref{kin}.

The velocity fields for NGC\,2787 are consistent with pure rotation in a disk oriented along position angle $PA\sim50/230^\circ$ east of north, with blueshifts to the southwest and redshifts to the northeast and a high projected velocity amplitude of $\sim250$\,\kms. 

NGC\,4450 also presents velocity fields indicating rotation in the disk of the galaxy, with blueshifts to the south and redshifts to the north of the nucleus, with a projected velocity amplitude of $\sim150$\,\kms. In addition to the rotation pattern, excess blueshifts of up to $-150$\,\kms are observed at 1\arc\ east of the nucleus, in a region comparable in size to the spatial resolution of our data.

\subsection{Velocity dispersion maps}


The \nii\ velocity dispersion map for each galaxy is presented in the mid-right panel of Figures~\ref{all-figs3982}--\ref{all-figs4450}, with black regions corresponding to masked locations due to bad fits. Similar maps for the \ha, \sii\ and \oi\ emission lines are shown in Figures  \ref{figs39}--\ref{figs44}.

NGC\,3982 shows $\sigma$ values ranging from $\sim50$\,\kms\ to $\sim130$\,\kms, with the highest values observed within 82\,pc from the nucleus and the smallest values in the partial circumnuclear star forming ring at 328 - 492\,pc ($r \approx$4--6\arc) from the nucleus. It can also be noticed that the forbidden lines show slightly larger $\sigma$ values than \ha, suggesting that they trace emission from kinetically ``hotter" gas.


The $\sigma$ values for NGC\,4501 are higher than $\sim120$\,\kms\ over a large part of the FoV, up to distances of 324\,pc  (4\arc) from the nucleus. Higher values of $\sim150$\,\kms\ are observed in a small patch (89.1\,pc$\times$153.9\,pc) at 121.5\,pc (1\farcs5) south of the nucleus. This region is co-spatial with excess redshifts observed in the velocity fields. On the other hand, very small $\sigma$ values ($\sim30-50\,$\kms) are observed at 486\,pc (6\arc) west of the nucleus, where another region of excess redshifts is observed in Figs.~\ref{all-figs4501} and \ref{figs45}. As for NGC\,3982, the forbidden lines show an average $\sigma$ value larger than those of \ha.

NGC\,2787 shows the highest $\sigma$ values of up to 150\,\kms\  within 63\,pc from the nucleus for all emission lines. Outside the nucleus there is an asymmetry in the distribution of $\sigma$ values: while to the south of the nucleus the lowest values of $\sim30-60$\,\kms\ are observed, to the northeast of the nucleus  $\sigma\ge120$\,\kms.




For NGC\,4450  the highest $\sigma$ values for all emission lines are observed within 81\,pc from the nucleus.
The \ha\ and \sii\ show $\sigma\sim100$\,\kms, while some higher values ($\sim200$\,\kms) are observed for the \nii\ at these locations. The smallest $\sigma$ values are $\sigma\sim50-70$\,\kms, observed in the spiral arms.

\subsection{Electron density}

The ratio of the fluxes \sii\,$\lambda$6716/$\lambda$6731 was used to obtain the electron density {\it N$_{e}$} using the {\sc temden} routine of the {\sc nebular} package from {\sc stsdas/iraf}, assuming an 
electron temperature for the ionized gas of 10\,000\,K. The bottom left panels of Figures~\ref{all-figs3982} - \ref{all-figs4450} show the gas electron density distribution for all galaxies.

The highest electron density values of about 3000\,cm$^{-3}$ are found within the inner 82\,pc (1$^{\prime\prime}$) of NGC\,3982. For the other galaxies the density values range from 600 to 1500\,cm$^{-3}$, and are similar to those
obtained in  similar studies of the inner kiloparsec of active galaxies \citep[e.g.][]{couto2013,lena15}

 \subsection{Stellar kinematics}

In order to obtain measurements for the Line-of-Sight Velocity Distribution (LOSVD) of the stars we used the  Penalized Pixel Fitting (pPXF)  routine of \citet{cappellari} to fit stellar absorptions in the spectral range from 5600 -- 6900\,\AA. The fitting of the galaxy spectra is done by using template spectra under the assumption that the LSVD of the stars is well reproduced by Gauss-Hermite series. We used selected simple stellar population synthetic spectra from the \citet{bruzual} models, which have similar spectral resolution to those of our GMOS data \citep[e.g.][]{allan14}.

The corresponding stellar velocity field and velocity dispersion ($\sigma_*$) maps for each galaxy are shown in the bottom-central and bottom-right panels of Figures~\ref{all-figs3982}--\ref{all-figs4450}.  White/black regions in the velocity/velocity dispersion maps correspond to masked locations were the uncertainties of the measurements are larger than 50 km\,s$^{-1}$.

The stellar velocity fields of all galaxies show signatures of rotation. For NGC\,3982, the stellar kinematic maps  are very noisy, but  the stellar velocity field shows a similar trend of rotating disk as observed for the gas, with blueshifts observed to the northeast of the nucleus and redshifts to southwest of it. We were able to measure the stellar kinematics only within the inner  2$^{\prime\prime}$ of NGC\,4501, which shows a clear rotation pattern with blueshifts observed to the northwest of the nucleus and redshifts to southeast of it, with a velocity amplitude of $\sim$150\,\kms. A similar velocity amplitude is observed for NGC\,2787 with blueshifts observed to the west and redshifts to the east, showing a similar signature of rotation as observed in the gas velocity field. The orientation of the line of nodes of the gas and stars seems to be misaligned by 30--40$^\circ$. However, we were not able to fit the stellar absorptions at most locations to the south of the nucleus of this galaxy, possible due to the larger extinction at this side, as seen in the structure map (top-left panel of Figure~\ref{all-figs2787}). NGC\,4450 shows a clear  rotating disk pattern with the line of nodes oriented along the north-south direction. The deviations from pure rotation seen in the gas velocity field are  not observed in the stellar velocity field of this galaxy.

The stellar velocity dispersion map of NGC\,3982 shows values smaller than 100\,\kms\ at most locations, being smaller than that observed for the gas at the same locations. NGC\,4501 shows $\sigma_*$ values overall larger than those observed for the \nii\ emitting gas and a partial low-$\sigma_*$ ring ($\sigma_*<100$\,\kms) is observed surrounding the nucleus at 1$^{\prime\prime}$. Similar structures have been observed for other active galaxies and attributed to being originated from intermediate age (100 Myr -- 2 Gyr) stellar populations \citep{mrk1066_pop,mrk1157_pop,llp_stel}.  For NGC\,2787, the $\sigma_*$ map shows values larger than 150~\kms\ at most locations, suggesting that although the stellar velocity field shows a clear rotation pattern, the stellar motions are dominated by random orbits, instead of ordered rotation in the plane of the disk. Finally, NGC\,4450 shows $\sigma_*$  values in the range from 70 to 130\,\kms, similar to that observed at extra-nuclear regions for the gas.  




\section{Rotation disk model}\label{rotmod}

As seen in Figs.~\ref{all-figs3982}--\ref{all-figs4450}, both the gas and the stellar velocity fields for all galaxies show a rotation pattern, with the gas presenting some deviations from pure rotation due to non-circular motions. Here we model only the gas velocity fields with a rotation model due to the fact that the stellar velocity fields are much noisier but present similar rotation patterns. We used a simple analytical model, assuming
that the gas has circular orbits at the plane of the galaxy  \citep{van,bertola}, as done in previous works by our group \citep[e.g.][]{couto2013,allan14,allan14b,lena15,lena16}.
The expression for the circular velocity is given by 

\[
V_{\rm mod}(R,\Psi)=\upsilon_{s}+ \]
\begin{equation}
~~~~~\frac{AR\cos(\Psi-\Psi_{0})\sin(i)\cos^{p}(i)}{\left\{R^2[\sin^2(\Psi-\Psi_{0})+\cos^2(i)\cos^2(\Psi-\Psi_{0})]+C_o^2\cos^2(i)\right\}^{p/2}},
\label{model-bertola}
\end{equation}
where {\it A} is the velocity amplitude, 
$\Psi_o$ is the position angle of the line of nodes,
C$_o$ is a concentration parameter, defined as is the radius where the rotation curve reaches 70\,\% of the velocity amplitude,
{\it i} is the disc inclination in relation to the plane of the sky ({\it i} = 0 for face-on disc),
{\it R} is the radial distance to the nucleus projected in the plane of the sky with the corresponding position angle $\Psi$,
and $\upsilon_{s}$ is the systemic velocity of the galaxy.
The parameter {\it p} measures the slope of the rotation curve where it flattens, in the outer region of the galaxy
and it is limited between 1 $\le$ {\it p} $\le$ 3/2. For {\it p}\,=\,1 the rotation curve at large radii is asymptotically flat
while for {\it p}\,=\,3/2 the system has a finite mass.

We used an IDL\footnote{\rm $http://www.harrisgeospatial.com/ProductsandSolutions/Geospatial$\\$Products/IDL.aspx$} based routine to fit the above equation to the observed \nii$\lambda6583$ velocity fields using the MPFITFUN routine \citep{mark09} to perform the non-linear least-squares fit, where initial guesses are given for each parameter and the routine returns their values for the best fitted model. As all lines show similar velocity fields, we have chosen the \nii\ kinematics to perform the fit, as the \nii$\lambda6583$ is the strongest line observed at most locations for all galaxies. During the fit, the location of the kinematical center was fixed to the position of the peak of the continuum emission and the parameter $p$ was fixed to $p=1.5$.

\begin{figure*}
\begin{center}
    \includegraphics[scale=0.5]{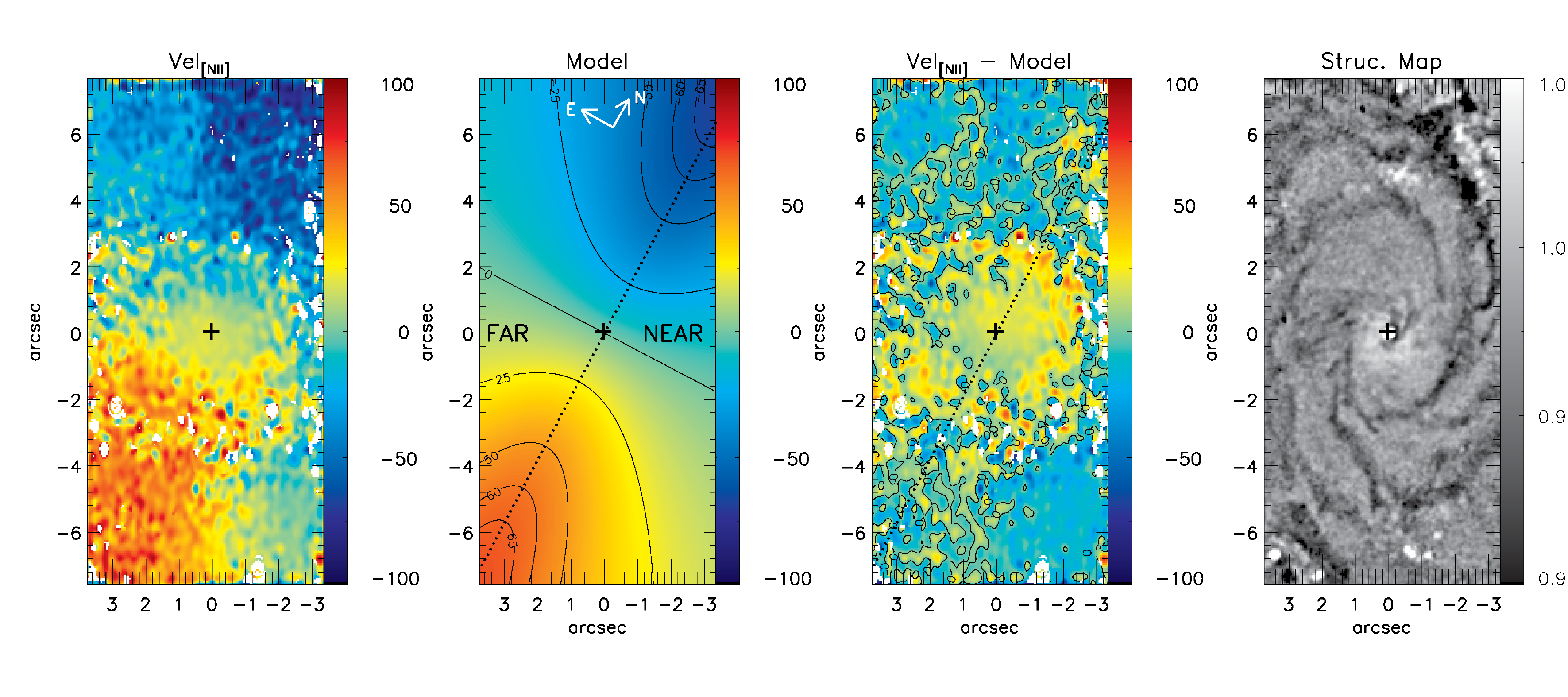} 
\caption{NGC\,3982: From left to right: \nii\ velocity field, rotating disc model for the [N\,{\sc ii}] velocity field (km\,s$^{-1}$), residual
map (between the observed, modeled velocities) and structure map. The dotted line displays the position
of the line of nodes, the structure map and the near and far sides of the galaxy are indicated.} 
\label{model39}
\end{center}
\end{figure*}

\begin{figure*} 
\begin{center}
 \includegraphics[scale=0.5]{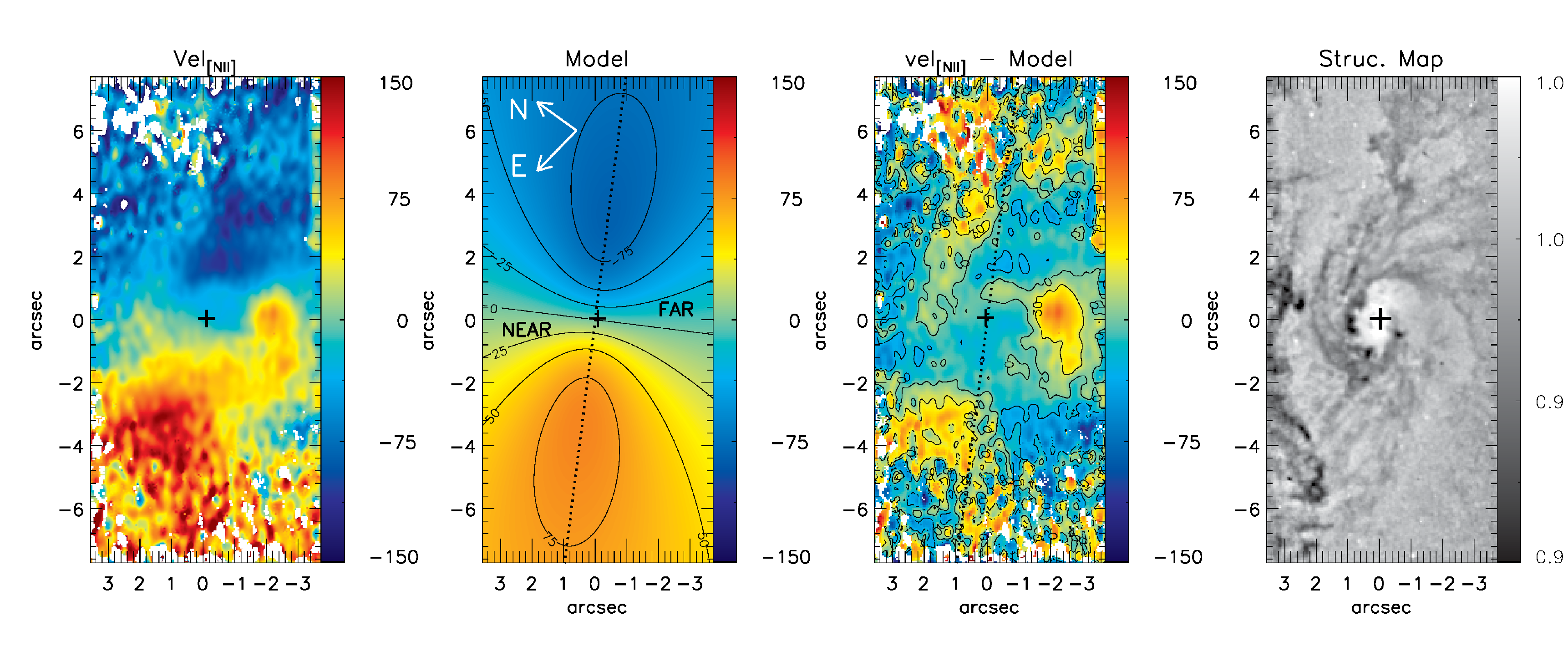} 
  \caption{Same as Fig.\,\ref{model39} for NGC\,4501} 
\label{model45}  
\end{center}
\end{figure*}

\begin{figure*} 
\begin{center}
 \includegraphics[scale=0.5]{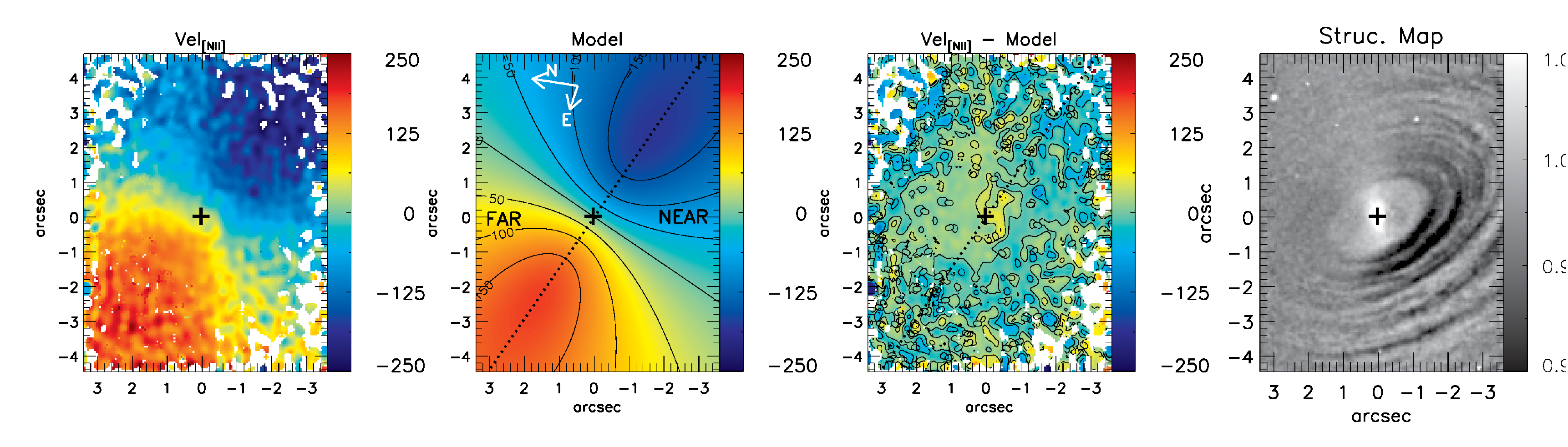} 
   \caption{Same as Fig.\,\ref{model39} for NGC\,2787} 
\label{model27}  
\end{center}
\end{figure*}

\begin{figure*} 
\begin{center}
 \includegraphics[scale=0.6]{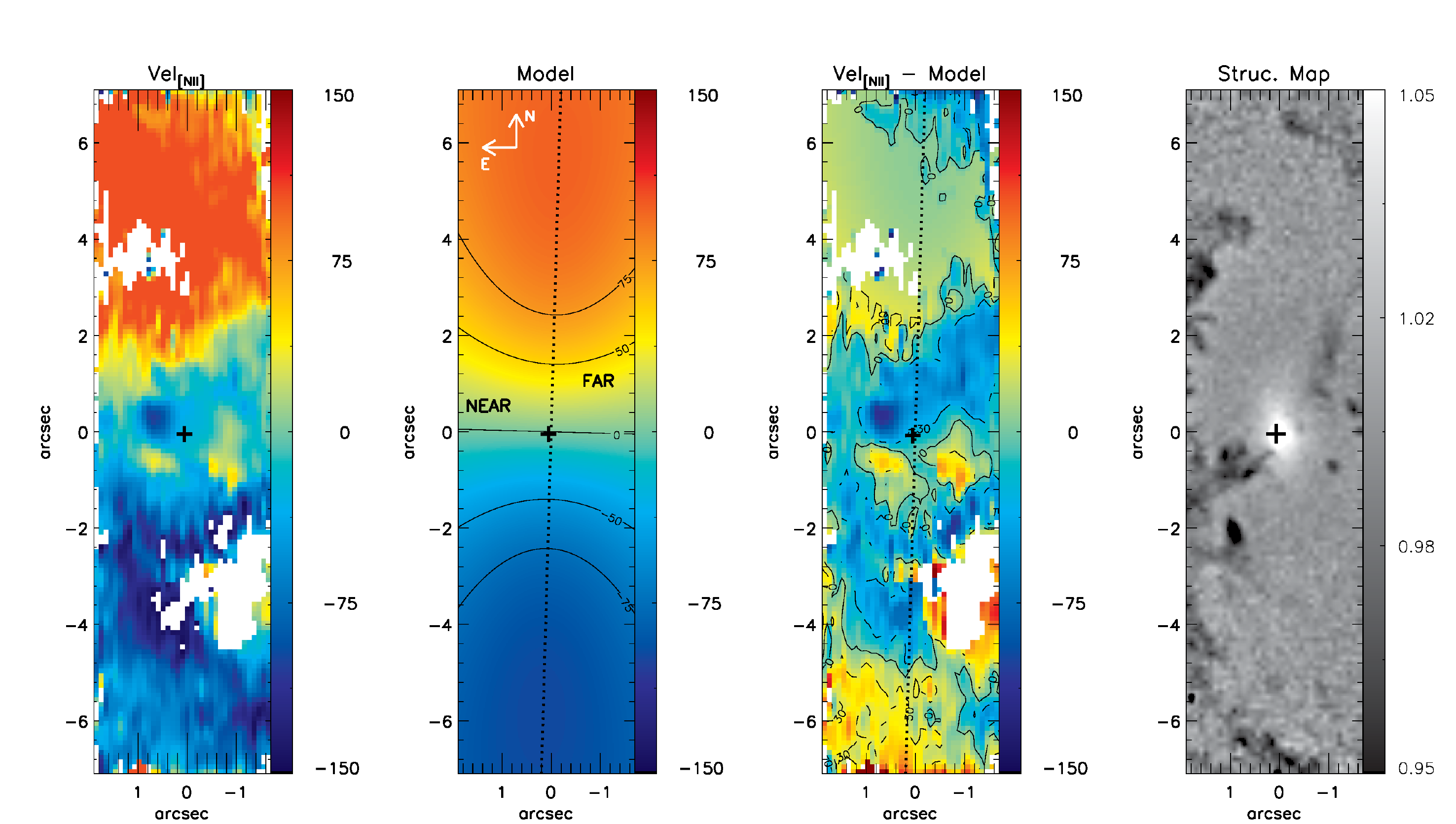} 
   \caption{Same as Fig.\,\ref{model39} for NGC\,4450} 
\label{model44}  
\end{center}
\end{figure*}

In Figures~\ref{model39}--\ref{model44} we show the \nii\ velocity field (left panel), resulting model (middle left panel), the residual map (middle right panel) and a structure map (right panel) for NGC\,3982, NGC\,4501, NGC\,2787 and NGC\,4450, respectively, while the resulting fitted parameters are shown in table~\ref{tab39}. The values of the disk inclination and the amplitude of the rotation curve are coupled, meaning that somewhat higher or lower inclinations give similar fits for corresponding somewhat smaller and larger amplitudes, respectively. A similar coupling is observed between $i$ and $c_0$ and these values should thus be considered with caution.

\begin{table*} 
\begin{center}
\caption{Parameters derived from our modeling for galaxies}
\vspace{0.5cm}
\begin{tabular}{c c c c c c}
\hline
\hline
Galaxy   &  A (km\,s$^{-1}$) &     $\upsilon_{s}$ (km\,s$^{-1}$)  & $\Psi_{0}$  & $c_{o}$   & $i$ \\
\hline
NGC\,3982 &  301 $\pm$ 5   & 1103\,$\pm$\,10 & 173$^{\circ}\,\pm$\,2$^{\circ}$ & 6.8\,$\pm$\,0.1$^{\prime\prime}$ & 72$^{\circ}$\,$\pm$\,2$^{\circ}$ \\
NGC\,4501 & 206\,$\pm$\,11 & 2244\,$\pm$\,12 & 133$^{\circ}\,\pm$\,3$^{\circ}$ & 2.0\,$\pm$\,0.1$^{\prime\prime}$ & 47$^{\circ}\,\pm$\,3$^{\circ}$ \\
NGC\,2787 & 585\,$\pm$\,4  & 615\,$\pm$\,11  & 76$^{\circ}\,\pm$\,1$^{\circ}$  & 2.1\,$\pm$\,0.1$^{\prime\prime}$ & 46.$^{\circ}\,\pm$\,2$^{\circ}$ \\
NGC\,4450 & 502\,$\pm$\,13 & 1910\,$\pm$\,10 & 192$^{\circ}\,\pm$\,4$^{\circ}$ & 6.5\,$\pm$\,0.3$^{\prime\prime}$ & 63.$^{\circ}\,\pm$\,3$^{\circ}$ \\
\hline
\end{tabular}
\label{tab39}
\end{center}
\end{table*}

\section{Discussion}\label{discussion}

In this section we present a small review of what is already known for each galaxy of our sample and discuss our results in comparison (or addition) to those of these previous works.

\subsection{NGC3982}

NGC\,3982 (or UGC\,6918) is part of the Ursa Major cluster. 
Optical and near-infrared broad band images reveal the presence of a small nuclear bar extending by 10\,\arc\  along $PA\sim30^{\circ}$ east of north, while at larger scales it displays a multi-armed spiral pattern, seen in a B-I color index map and \ha\ narrow band images \citep{knapen02,regan99,perez-ramirez00}. Multi-spiral arms are also seen at small scales, as revealed by the HST F606W image \citep{slopes,munoz-marin07}. At radio wavelengths,  a weak elongated feature of 4~kpc length is observed along the north-south direction in 6~cm, plus unresolved nuclear emission, probably due to the AGN \citep{ho01}. The radiation field of the AGN seems also to heat the dust  close to the nucleus, as revealed by a very red near-IR color \citep{perez-ramirez00}. In addition, the mass of molecular gas is larger than the stellar mass within the inner $1^{\prime} \times 1^{\prime}$ \citet{martinsson13b}.

The WHAN diagram of NGC\,3982 (Fig.~\ref{all-figs3982}) confirms the signature of a strong AGN (sAGN) at the nucleus within the inner arcsec (82pc) in agreement with the Seyfert 2 classification \citep[e.g.][]{ho97,quillen,veron,trippe}. This region also presents a higher velocity dispersion ($\sigma \geq$ 130 \kms) than its surroundings ($\sigma \leq$ 80 \kms)-- see Fig.~\ref{all-figs3982}, which could suggest the presence of a mild unresolved nuclear outflow. 
 In addition, strong AGN signatures are also observed at about 328\, pc (4$^{\prime\prime}$) from the nucleus, at locations that surround star forming regions in the ring. We attribute this signature to 
an underlying young/intermediate age stellar population with strong H$\alpha$ absorption leading to an underestimate of the \ha\ flux and thus to a large \nii/\ha\ line ratio. This could also
explain the weak AGN signatures at the surrounding regions, where the \ha\ fluxes are weak. Another possibility to explain this emission, could be the presence of post-AGB stars as already claimed to explain the LIER emission in large samples of galaxies \citep{belfiore16}.  The star forming regions identified in the flux maps and already known \citep[e.g.][]{buta,comeron} are properly classified as ``Starburst" by the WHAN diagram, although some ``noise" is observed. 

Large scale IFU observations show that the stellar and gas velocity fields are consistent with regular rotation \citep[e.g.][]{westfall11,martinsson13}, while our small scale gas kinematics suggest the presence of deviations from pure rotation, as observed in Fig.~\ref{model39}.  By comparing the fitted parameters from the rotation disk model (Table~\ref{tab39}) with those found in the literature, we observe a large discrepancy among the values for the large scale disk and ours. The major axis of the galaxy is oriented along $\Psi_0=25.1^\circ$, as quoted in the Hyperleda database \citep{paturel03}.  \citet{ciesla} used observations at 8$\mu$m and 500$\mu$m  with the Herschel telescope and found that the major axis of the mid infrared emission is oriented along $\Psi_0=138^{\circ}$ which is 40 degrees smaller than the 173$\pm$2  derived from our internal gas kinematics, while \citet{martinsson13} present \oiii\ and stellar kinematics measured for an 1$^\prime$ FoV from  PPak integral field spectroscopy and found $\Psi_0=191.6^\circ\pm0.5^\circ$, which is about 20$^\circ$ larger than ours. These discrepancies are probably due to the distinct FoV of the observations and the complex gas motions present within the inner few hundred parsecs of the galaxy. The systemic velocity of the galaxy obtained here is in good agreement with that of \citet{martinsson13} and quoted in the Hyperleda database.

\subsection{NGC4501}

NGC\,4501 belongs to the Virgo cluster.
High-resolution interferometer observations of the $^{12}$CO ($J=1-0$) emission in the central 5~kpc using the Nobeyama Millimeter Array show two molecular gas structures: (i) a nuclear concentration within the inner 5\arc\ with mass of $1.3\times10^8$\,M$_\odot$, showing non-circular motions and (ii) large scale spiral arms, along which molecular gas streaming motions towards the center are observed \citep{onodera}. Multiple large scale spiral arms and two small scale arms are observed in the B-band \citep{elmegreen87} and K-band \citep{elmegreen99} images. WFPC2 F606W HST images reveal nuclear dust spirals \citep{carollo98}.

\citet{mazzalay13} used near-infrared integral field spectroscopy of the inner 3\arc$\times$3\arc\ of NGC\,4501 obtained with the SINFONI instrument on the Very Large Telescope (VLT) to map the
H$_2$\,2.12\,$\mu$\,m emission. They found that the H$_2$ flux distribution shows two components: an asymmetric nuclear component surrounded by two arcs, one to the northwest and the other to the southeast of the nucleus, that seem to be correlated with nuclear dust lanes seen in an WFPC2 F547M HST image and located inside edges of two peaks seen in CO image from \citet{onodera}, indicating that the warm molecular gas traced by the H$_2$ line emission
is closer to the nucleus than the cold molecular gas. In addition, \citet{mazzalay13} report than the spectra of NGC\,4501 show no sign of ionized gas emission (e.g. Br$\gamma$). The stellar kinematics of the inner 3\arc$\times$3\arc\ shows regular rotation with the line of nodes oriented along $PA=140^\circ$ east of north, while the H$_2$ kinematics shows additional non-circular motions \citep{mazzalay14}. The authors analyze the H$_2$ kinematics based on a residual map, constructed by subtracting a rotating disk model from the observed velocity field. This map shows two main structures: one observed in blueshift to the northwest of the nucleus, co-spatial with a dust spiral arm, and another to the southwest of the nucleus seen in redshift. \citet{mazzalay14} interpret these structures as being due to outflows from the nucleus.

Recently, \citet{repetto} used the same GMOS-IFU data of NGC\,4501 used in this work (from Gemini archive) to study the gas kinematics and stellar populations. However, their analysis is presented only for the inner 6\farcs4$\times$5\farcs4 nuclear region. The velocity field and FWHM map for the [N\,{\sc ii}]$\lambda6583$ emission line shown in \citet{repetto} is consistent with ours, although their maps are much ``noisier" than ours. \citet{repetto} found that old stellar populations dominate the continuum emission from the inner region of the galaxy. They conclude that the gas kinematics is dominated by non-circular motions 
and reproduced by an exponential disk model, with a maximum expansion velocity of 25~km\,s$^{-1}$ and major axis along $\Psi_0\approx 137^\circ$. The authors argue also that the kinematics of the NaD$\lambda5892$ is consistent with outflowing material from the center of NGC\,4501 along two inner pseudo-spiral arms.

Our orientation for the line of nodes of the gas velocity field is also in good agreement with that quoted in the Hyperleda database ($\Psi_0=138^\circ$) for the large scale disc, as well as with those obtained for the central region from stellar kinematics of the inner $3^{\prime\prime}\times3^{\prime\prime}$ \citep[$\Psi_0=140^\circ$--][]{mazzalay14}  and H$\alpha$ velocity field of the inner  6\farcs4$\times$5\farcs4  \citep[$\Psi_0=137^\circ$--][]{repetto}, while the systemic velocity is about 25\,\kms\ smaller than that obtained from the optical measurements and quoted in Hyperleda database.

NGC\,4501 is cataloged as harboring a Seyfert 2 nucleus \citep{veron}, while according to our WHAN diagram its nuclear emission corresponds to a "weak AGN". X-ray emission from the nucleus of this galaxy gives a low luminosity of L$_{2--10 KeV} \approx 3.5 \times 10^{39}$\,erg s$^{-1}$ \citep{liu11}. A similar luminosity is also observed for the \oiii$\lambda5007$ emission line, $L_{OIII} \approx 9.6\times 10^{39}$\,erg s$^{-1}$ \citep{brightman08}, while strong AGNs usually have larger \oiii\ luminosities. 
In addition, the nucleus of NGC\,4501 falls at the region between Seyferts and LINERs in the \oiii$\lambda5007$/H$\beta$ vs. \nii$\lambda6583$/\ha\ diagnostic, as shown in \citep{brightman08}. Thus, our observations suggest that NGC\,4501 nuclear activity is better classified as LINER, instead of Seyfert\,2.

An intriguing feature is seen at $\sim$6\arc\ west of the nucleus, where the WHAN diagram shows AGN values (very close to the corner -- in the WHAN diagram -- that separates Starburts, strong and weak AGNs). We attribute this apparent sAGN excitation as being due to the presence of a young stellar cluster, as indicated by a slightly enhancement in the \ha\ flux map of Fig.~\ref{figs45}, and possible shocks due to supernovae explosions, that would enhance the \nii/\ha\ line ratio \citep[e.g.][]{sutherland93,viegas94,thaisa07}. Assuming the \ha\ emission is originated by gas photo-ionized by young stars, and using the photo-ionization models from \citet{dors08} and the observed \ha\ equivalent width for position A, we estimate an age of $>10$\,Myr, which is consistent with the presence of evolved stars.

\subsection{NGC2787}


Using long-slit spectra obtained with HST STIS at parsec-scale resolution, \citet{sarzi} derived the mass of the SMBH as being about 10$^{8}$\,M$_{\odot}$, by modeling the gas kinematics.  HST and ground based broad band images show a very complex morphology comprising a large inner disk, a nuclear bar oriented along $PA\sim-20^\circ$  and an off-plane dust disk in the central regions \citep[e.g.][]{sarzi,erwin03,erwin03b}. 
Inside the bar, the inner disk is tilted relative to the orientation of the stellar distribution
\citep{erwin2002}, while the large scale H\,{\sc i} distribution is also  found to be misaligned relative to the optical emission, suggesting the presence of a dark halo \citep{shostak87}. Chandra 0.5--1.5~keV observations show unresolved nuclear emission consistent with a stellar origin, with only a small contribution from hot gas \citep{li11}. So far, there are no studies available about the gas and stellar kinematics of the central region of NGC\,2787.

NGC\,2787 shows a WHAN diagram (Fig.~\ref{all-figs2787}) with wAGN signature observed at the nucleus and RG signatures else where, confirming that its nuclear activity is of the LINER type \citep[]{veron}, with a weak broad H$\alpha$ component observed at the nuclear spectrum (Fig.~\ref{maps}), already previously detected \citep{ho97}.

The systemic velocity for NGC\,2787 derived in Sec.~\ref{rotmod} is in reasonable agreement with that presented at Hyperleda database ($\upsilon_s=606\pm40$\,\kms), while the orientation of the line of nodes that we have obtained is 37$^\circ$ smaller than the PA of the major axis of the large scale disk listed in Hyperleda. On the other hand, it is known that NGC2787 present a complex structure in the central region \citep[e.g.][]{erwin03,erwin03b} and thus a misalignment between the large and small scale disk can be expected. Indeed, the $PA$ of the line of nodes we have derived for the circumnuclear gas kinematics of 100$^\circ$ is consistent with the orientation of the nuclear bar in HST images \citep{erwin03b}.

Another feature observed in the gas velocity field is an increase in the velocity dispersion at the nucleus extending to $\approx$2$^{\prime\prime}$ (126 pc) to the south of the nucleus, which could indicate a mild AGN outflow.

\subsection{NGC4450}
NGC\,4450 is an anemic spiral galaxy in the Virgo cluster.
Its nuclear spectrum shows a weak broad double-peaked \ha\ profile, interpreted as the signature of the outer parts of a  relativistic accretion disk \citep{ho97}. 
Fabry-Perot observations at a seeing of $\sim$1.5$^{\prime\prime}$ and a field of 1.8$^\prime$ reveal a patchy \ha\ distribution and a perturbed velocity field with the orientation of the line of nodes along $PA=351^\circ\pm9^\circ$ east of north and with a steep velocity gradient of 200\kms\ around the nucleus \citep{chemin06}. Similar distribution is observed in H\,{\sc i} emission \citep{cayatte90}. Interferometric observations show only weak CO emission and a total mass of cold gas of $\sim10^9\,$M$_\odot$ is observed for the whole galaxy \citep{helfer03}.
The nuclear region of NGC\,4450 shows two long dusty spirals in the main disk along with some flocculent structures, but with no star formation associated to the dusty spirals \citep{elmegreen}.

\citet{cortes} present optical IFU observations of the inner 20\arc$\times$40\arc\ of NGC\,4450. They found that the stellar velocity field is dominated by rotation, but very perturbed, with the orientation of the line of nodes changing from 175$^\circ$ at the center to 160$^\circ$ at 15-25\arc\ of the nucleus. The \oiii\ velocity field is misaligned relative to the stellar kinematics, with the orientation of the line of nodes along $PA\sim190^\circ$ east of north. \citet{cortes} interpret this misalignment as being due to non-circular motions or due to emission of gas located in a tilted gas disk relative to the stellar disk, produced by an accretion event or minor merger.

The orientation of the line of nodes  derived in Sec.~\ref{rotmod} ($\Psi_0=192^\circ$) is in good agreement with that of the large scale disk \citep{dicaire}, and consistent with the orientation of the line of nodes observed for the \oiii\ emission at the inner 20\arc\ \citep{cortes}. 
 A knot of residual blueshift (after the subtraction of the rotation model) at $\approx$ 1$^{\prime\prime}$ (81 pc) to the east of the nucleus (the near side of the galaxy) and residual redshift at a similar distance to the west (the far side of the galaxy) possibly indicate the presence of a nuclear outflow in the east-west direction (direction of the largest extent of the [NII] flux map). However it should be noticed that the size of the structures seen in blueshifts and redshifts in the residual maps are comparable to the spatial resolution of our data.  Another feature of the gas kinematics that our measurements revealed is an increase in the [NII] velocity dispersion at the nucleus,  which could be due to a previous plasma ejection related to the compact nuclear outflow. Alternatively, the larger velocity dispersion at the nucleus could also be due to unresolved rotation.

Our WHAN diagram for  NGC\,4450  shows strong AGN features in the inner 0\farcs5, surrounded by a ring of weak AGN signature. NGC\,4450 was previously cataloged as harboring a LINER nucleus \citep{veron,kewley06}, but it presents stronger X-ray emission than NGC\,4501, originally classified as Seyfert and with a WHAN diagram indicating that Seyfert is better classification for its nuclear emission. \citet{liu11} presents  a 0.3-8 keV flux of $1.19\times10^{-12}$\,erg s$^{-1}$, which corresponds to a luminosity of  $L_{0.3-8\,\rm KeV} \approx 3.6 \times 10^{40}$\,erg s$^{-1}$, that is one order of magnitude larger than the values observed for NGC\,2787 and NGC\,4501. On the other hand, NGC\,4450 shows a low \oiii$\lambda5007$ luminosity \citep[$L_{OIII} \approx 6\times 10^{38}$\,erg s$^{-1}$,][]{balmaverde13}, smaller than commonly observed in sAGNs.

\subsection{Residual gas velocity maps vs. nuclear spirals}\label{kin}

The velocity residual maps (obtained as the difference between the observed velocity fields and the rotation model)  for NGC\,4501 and NGC\,4450 show that many of the kinematic structures of these maps are spatially correlated with the dust features seen in the structure map. The residual maps are shown in the central panels of Figs.~\ref{model39}--\ref{model44}. 
The residual map for NGC\,3982 shows some  structures revealing the presence of non-circular motions, although only a few of these structures are correlated with dust features.
For NGC 2787 (Fig.~\ref{model27}) the velocity residuals are small at all locations and no systematic structures are seen in the residual map, indicating that the adopted model is a good representation of the observed velocity field, which is dominated by the rotating disk component. 


In order to better analyze the structures in the velocity residual maps, we assume that the spiral arms observed in the large scale images are trailing to determine the near and far side of the disk,identified in the central panels of Figs. 6--9 that show the rotation model for each galaxy. We also show in these figures the structure maps at the same scale as the kinematic maps in order to verify possible correlations between dust and kinematic structures in the residual maps. This is motivated by previous results from our group in which we have found an association between gas inflows and nuclear spiral arms \citep[e.g.][]{fathi06,thaisa07,riffel2008,mrk79,allan14}.

The residual map for NGC\,3982 shows that the highest blueshifts of up to $\sim-50$\,\kms\ are observed to the N-NE of the nucleus in the far side of the galaxy, in a region where the structure map shows a strong dust spiral arm.
In addition, some redshifts are also observed associated to the inner part of the same spiral arm, but in the near side of the galaxy. A possible interpretation to this residuals is that they represent inflows of gas towards the nucleus. However, this interpretation should be taken with caution, as residuals of the order of 20--40\,\kms\ are observed also at other locations of the galaxy.  We can thus only state with certainty that the velocity residuals are correlated with the dust structures seen in the nuclear region.

A correlation between the velocity residuals and the dust structures is also observed for NGC\,4501, as can be seen by comparing the central-right and right panels of Fig.~\ref{model45}. 
Besides these correlations, a ``redshifted blob" is observed at 1--3\arc\ SW of the nucleus. A similar structure was  observed by \citet{mazzalay14} in a residual map for the H$_2$ kinematics, and interpreted as due to an outflow, together with an arc-shaped blueshifted outflow in the near side of the galaxy. These kinematic structures are also supported by \citet{repetto}. The outflow interpretation for the redshifted blob is supported, in our observations, by an increase in the gas velocity dispersion at the location of the blob, as seen in Fig.~\ref{all-figs4501} and Fig.~\ref{figs45}. In addition, some blueshifts to the west (in the far side of the galaxy)  and redshifts to the east (in the near side of the galaxy)  could be attributed to inflows towards the nucleus, but there are also similar residuals at other locations and again we can only state that these residuals are associated with the dust structures.

Finally, for NGC4450, the residual map shows blueshifts of up to --100 \kms, as well as similarly high redshifts. A disturbed kinematics for the gas in the central region of this galaxy has already been claimed by \citet{cortes}, who showed that the \oiii\ kinematics was misaligned relative to the stellar kinematics. Besides the usual correlation between the residuals and the dust structures, blueshifts of up to 150~\kms\ at 0\farcs5 NE of the nucleus and some redshifts observed at similar distance to the SW could be interpreted as due to a bi-conical outflow, although this is only a speculation, as other similar residuals are observed at other locations. The presence of a nuclear outflow is also supported by the increased velocity dispersion observed within the inner 1$^{\prime\prime}$ (Fig. \ref{all-figs4450}).



In summary, the gas kinematics, although dominated by rotation, shows deviations correlated with dust structures. Such structures usually trace shocks in the gas and we speculate that we are probing these shocks that may lead to loss of angular momentum allowing for the gas to move inwards to feed the AGN at the nuclei of the galaxies.

\section{Conclusions}

We have mapped the ionized gas kinematics and flux distributions in the central kiloparsec of NGC\,3982, NGC\,4501, NGC\,2787 and NGC\,4450 using GMOS IFS at a velocity resolution of $\sim$~120\,\kms\ and spatial resolution in the range 50--70 pc.  The four galaxies show extended emission for \ha, \nii\ and \sii\ emission lines, while \oi\ extended emission was observed only for NGC\,3982, to up to 2\arc\ from the nucleus. The main conclusions of this work are:

\begin{itemize}
\item The velocity field of all galaxies are dominated by rotation and reproduced by a disk model, under the assumption that the gas rotates at the plane of the galaxy at circular orbits. 

\item Besides the rotating disk component, the gas in NGC\,3982, NGC\,4501 and NGC\,4450 show non-circular motions, evidenced in the residual (observed velocity -- rotation mode) velocity maps. At least for the latter  two, these residuals are associated with dust features revealed in the structure maps.

\item The velocity residual map for NGC4501 reveals also a redshifted blob in the far side of the galaxy at 1--2~\arc\ SW of the nucleus that can be interpreted as due to a nuclear outflow. Possible outflows are also observed in NGC4450 as blueshifts and redshifts within the inner 1\arc\ to the northeast and southwest, respectively.


\item NGC\,2787 shows a very regular rotation with the orientation of the line of nodes misaligned by $\sim40^\circ$ relative to the large scale disk. This galaxy is known to show a complex morphology at the central region and the $PA$ of the line of nodes is consistent with the orientation of a nuclear bar.

\item The \ha\ equivalent width ($W_{H\alpha}$) vs. [N\,{\sc ii}]/H$\alpha$ (WHAN) diagrams show a wide range of values, with the nuclear emission of NGC\,3982 and NGC\,4450 showing a Seyfert signature, while for NGC2787 and NGC4101 a LINER signature is obtained.

\item NGC\,3982 shows a clear circumnuclear star-formation ring surrounding the nucleus at 4-6\,\arc, as seen in the flux maps and in the WHAN diagram.

\item A star forming region was detected at 6\arc\ west of the nucleus of NGC\,4501. The WHAN diagram shows values typical of Seyfert galaxies for this region and we interpret it as being originated by emission enhanced due to shocks from supernovae  explosions.

\item The excitation maps show that the AGN emission is very compact for all galaxies, being unresolved for NGC\,4501, NGC\,4450 and NGC\,4450. 

\end{itemize}

\section*{Acknowledgements}
 This work is based on observations obtained at the Gemini Observatory, 
which is operated by the Association of Universities for Research in Astronomy, Inc., under a cooperative agreement with the 
NSF on behalf of the Gemini partnership: the National Science Foundation (United States), the Science and Technology 
Facilities Council (United Kingdom), the National Research Council (Canada), CONICYT (Chile), the Australian Research 
Council (Australia), Minist\'erio da Ci\^encia e Tecnologia (Brazil) and south-eastCYT (Argentina).  C.B thanks to CNPq for financial support. 
R.A.R. acknowledges support from FAPERGS (project N0. 2366-2551/14-0) and CNPq (project N0. 470090/2013-8 and 302683/2013-5).

\appendix
\section{Flux distributions and kinematics}
\begin{figure*} 
\begin{center}
 \includegraphics[scale=0.7]{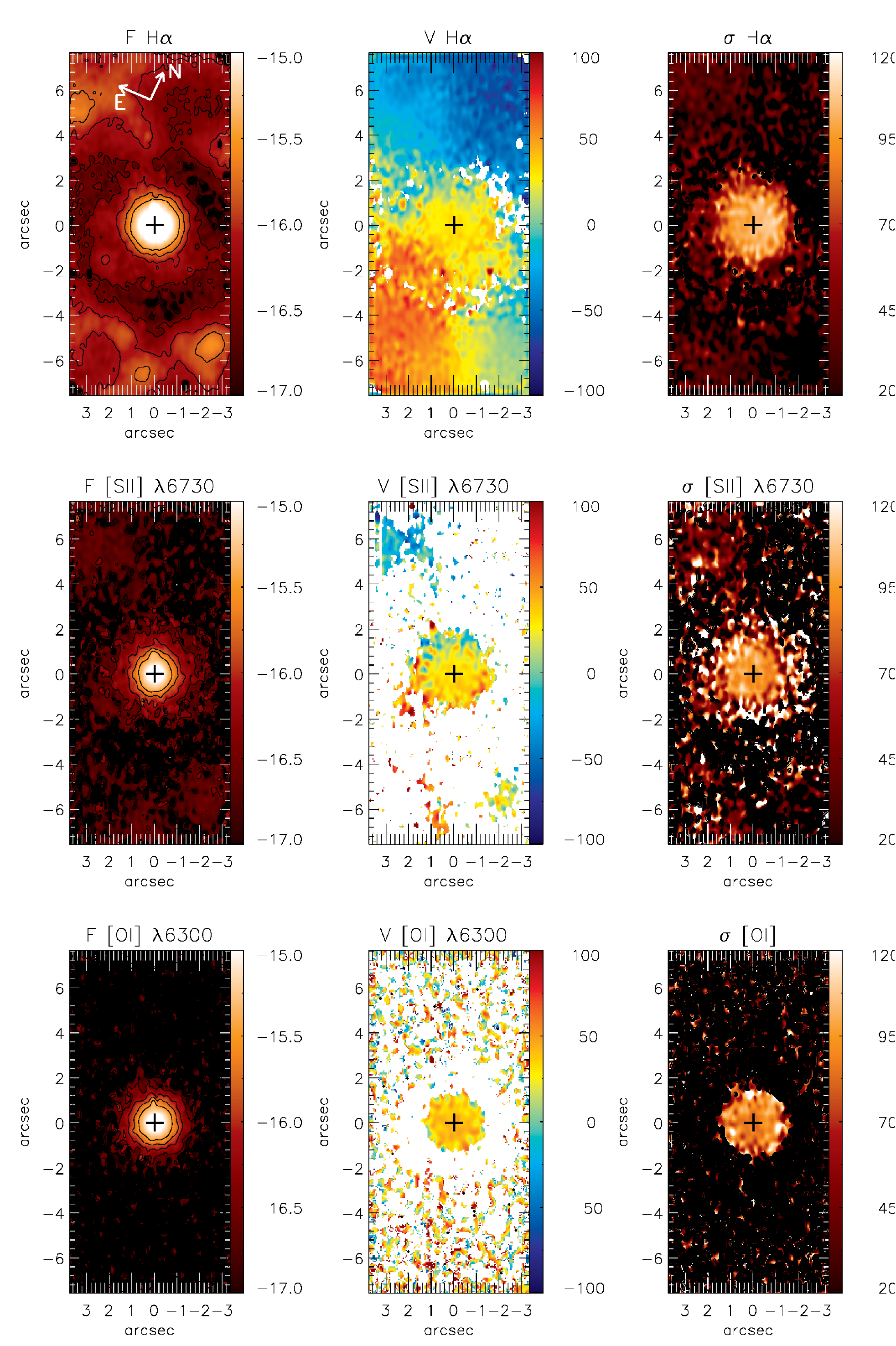} 
  \caption{Flux distributions for H$\alpha$, [S\,{\sc ii}]\,$\lambda$6730 and \oi$\lambda$6300  emission lines. The color bars show the flux scale in units of $10^{-17}$erg s$^{-1}$cm$^{-2}$ and the central cross marks the position of the nucleus. Black regions car masked regions and corresponds to locations where  no good fits of the line profiles could be obtained. Centroid velocity fields for the  \ha, \sii and \oi emitting gas. The central crosses mark the position of the nucleus and color bars  show the observed velocities in units of \kms\ relative to the systemic velocity of the galaxy. White regions correspond to locations where the lines where not detected or non good fits were possible.
Velocity dispersion ($\sigma$) maps for \ha, \sii and \oi emission lines for NGC\,3982. The central crosses mark the position of the nucleus and he color bars show the $\sigma$ values in \kms. Black regions correspond to locations where the lines where not detected or non good fits were possible.} 
\label{figs39}  
\end{center}
\end{figure*}

\begin{figure*} 
\begin{center}
 \includegraphics[scale=0.7]{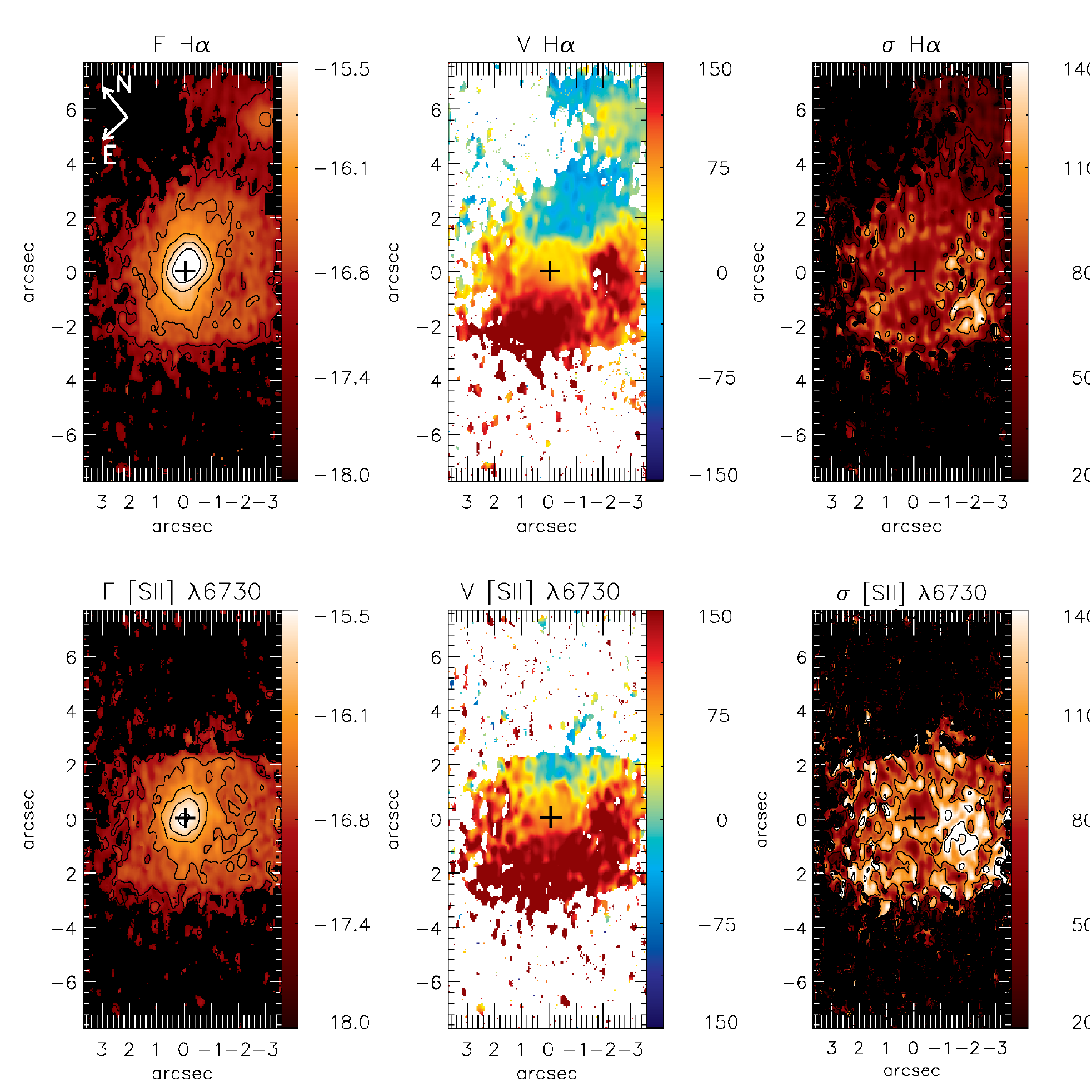} 
  \caption{Same as Fig.\,\ref{figs39} for NGC\,4501} 
\label{figs45}  
\end{center}
\end{figure*}

\begin{figure*} 
\begin{center}
 \includegraphics[scale=0.5]{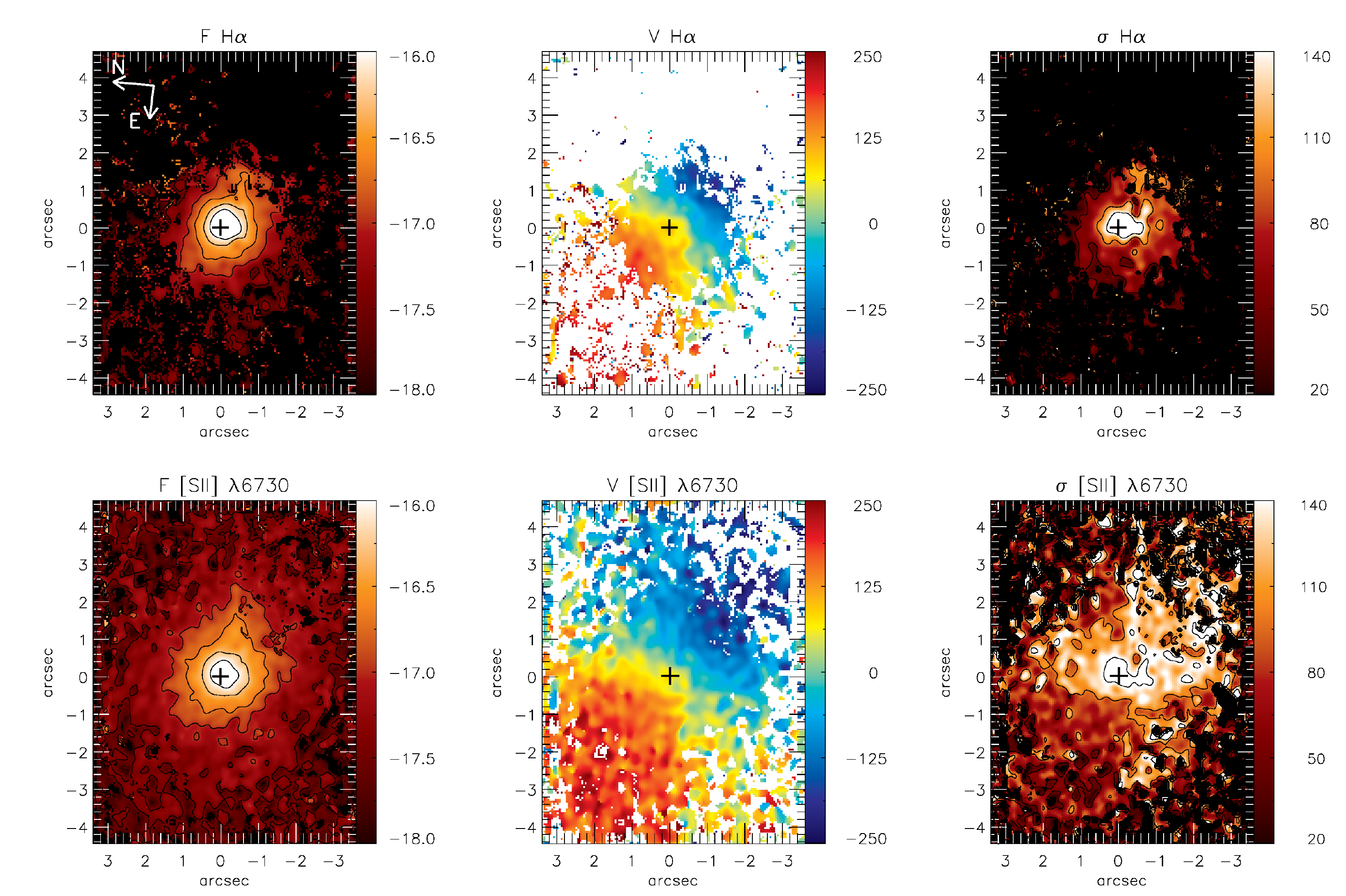} 
  \caption{Same as Fig.\,\ref{figs39} for NGC\,2787} 
\label{figs27}  
\end{center}
\end{figure*}

\begin{figure*} 
\begin{center}
 \includegraphics[scale=0.7]{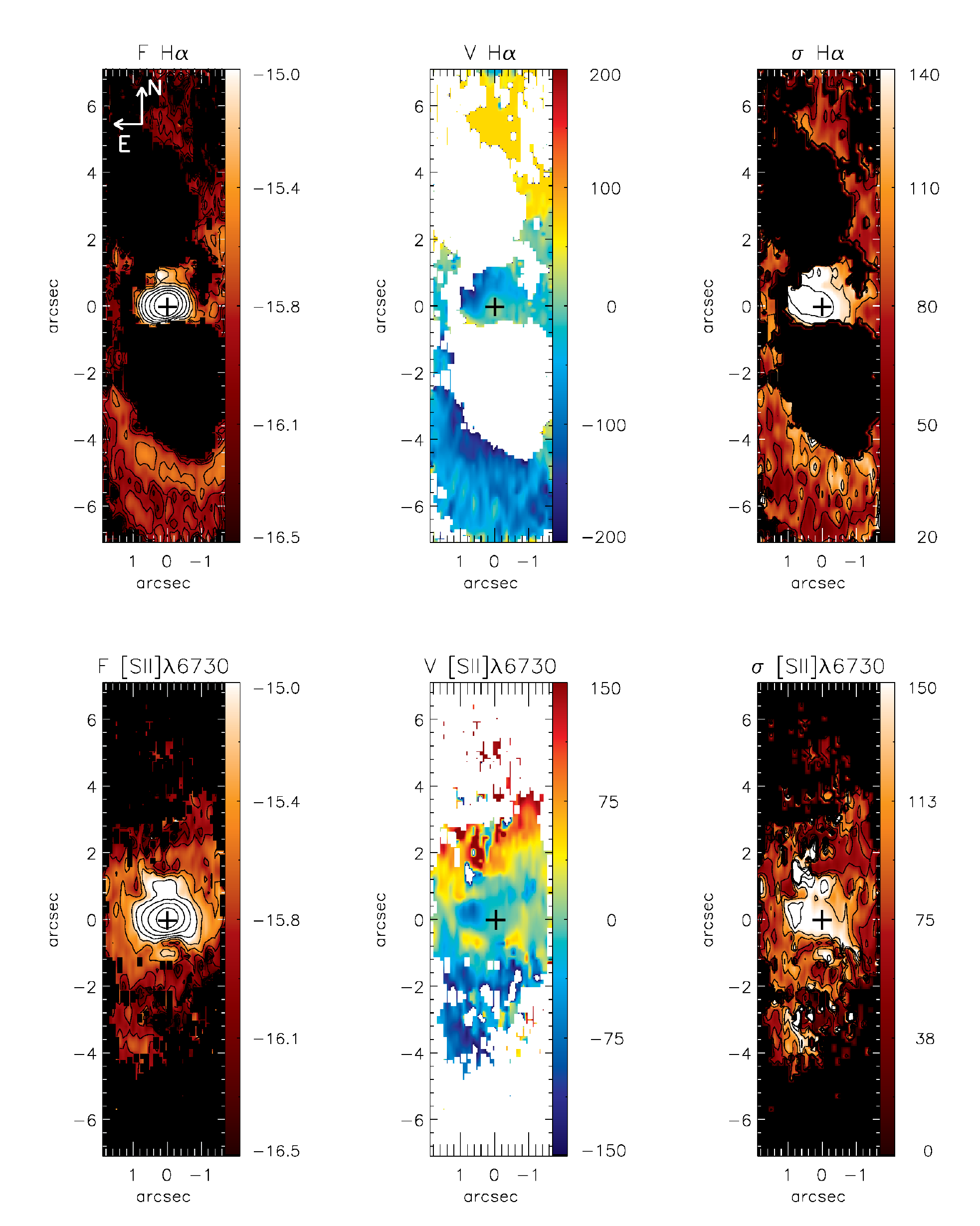} 
  \caption{Same as Fig.\,\ref{figs39} for NGC\,4450} 
\label{figs44}  
\end{center}
\end{figure*}

Figures~\ref{figs39} to \ref{figs44} show maps for the flux distributions, centroid velocity and velocity dispersion of the \ha\, [S\,{\sc ii}]\,$\lambda$6730 and \oi$\lambda$6300  emission lines for NGC\,3982, NGC\,4501, NGC\,2787 and NGC\,4450.

\label{lastpage}


\begin{thebibliography}{99}

\bibitem[\protect\citeauthoryear{Allen, Dopita \& Tsvetanov}{1998}]{allen98} Allen, M., G., Dopita, M., A., Tsvetanov, Z. 1998, AJ, 493, 571.

\bibitem[\protect\citeauthoryear{Allington-Smith et al.}{2002}]{allington-smith02} Allington-Smith, J., Graham, M., Content, R., Dodsworth, G., Davies, R., Miller, B. W., Jorgensen, I., Hook, I., Crampton, D., Murowinski, R., 2002, PASP, 114, 892.

\bibitem[\protect\citeauthoryear{Baillard et al.}{2011}]{baillard11} Baillard, A., et al. 2011, A\&A, 532, A74.

\bibitem[\protect\citeauthoryear{Balmaverde \& Capetti}{2013}]{balmaverde13} Balmaverde, B., Capetti, A., 2013, A\&A, 549, 114.	

\bibitem[\protect\citeauthoryear{Baldwin,  Phillips \& Terlevich}{1981}]{bpt} Baldwin, J. A., Phillips, M. M., Terlevich, R., 1981, PASP, 93, 5.

\bibitem[\protect\citeauthoryear{Belfiore et al.}{2016}]{belfiore16} Belfiore, F., et al. 2016, MNRAS, 461, 3111.

\bibitem[\protect\citeauthoryear{Bertola et al.}{1991}]{bertola} Bertola, F., Bettoni, D., Danziger, J., 1991, ApJ, 373, 369.

\bibitem[\protect\citeauthoryear{Brightman \& Nandra}{2008}]{brightman08} Brightman, M., Nandra, K., 2008, MNRAS, 390, 1241.

\bibitem[\protect\citeauthoryear{Bruzual \& Charlot}{2003}]{bruzual} Bruzual, G. \& Charlot, S. 2003, MNRAS, 344, 1000

\bibitem[\protect\citeauthoryear{Caldwell et al.}{1999}]{caldwell} Caldwell, N., Rose, J. A., \& Dendy, K. 1999, AJ, 117, 140

\bibitem[\protect\citeauthoryear{Capelo \& Dotti}{2017}]{capelo17} Capelo, P. R. \& Dotti, M., 2017, MNRAS, 465, 2643.

\bibitem[\protect\citeauthoryear{Cappellari \& Emsellem}{2004}]{cappellari} Cappellari, M., Emsellem, E. 2004, PASP, 116, 138.

\bibitem[\protect\citeauthoryear{Cayatte et al.}{1990}]{cayatte90} Cayatte, V., van Gorkom, J. H., Balkowski, C., Kotanyi, C., 1990, AJ, 100, 604.

\bibitem[\protect\citeauthoryear{Carollo et al.}{1998}]{carollo98} Carollo C. M., Stiavelli M., Mack J., 1998, AJ, 116, 68

\bibitem[\protect\citeauthoryear{Cid Fernandes et al.}{2010}]{cid10} Cid Fernandes, R., Stasi{\'n}ska G., Schlickmann M. S., Mateus A., Vale Asari N., Schoenell W., Sodr{\'e} L., 2010, MNRAS, 403, 1687

\bibitem[\protect\citeauthoryear{Cid Fernandes et al.}{2011}]{cidf} Cid Fernandes, R., Stasi{\'n}ska G., Mateus A., Vale Asari, 2011, MNRAS, 413, 1036

\bibitem[\protect\citeauthoryear{Ciesla et al.}{2014}]{ciesla} Ciesla, L., Boquien, M., Boselli, A., Buat, V., Cortese, L., Bendo, G. .J., Heinis, S., Galametz, M., Eales, S., Smith, M. W. L., Baes, M., Bianchi, S., de Looze, I., di Serego Alighieri, S., Galliano, F., et al. 2014, A\&A, 565, 128

\bibitem[\protect\citeauthoryear{Chemin et al.}{2006}]{chemin06} Chemin, L., Balkowski, C., Cayatte, V., Carignan, C., Amram, P., Garrido, O., Hernandez, O., Marcelin, M., Adami, C., Boselli, A., Boulesteix, J., 2006, MNRAS, 812, 857.

\bibitem[\protect\citeauthoryear{Colina et al.}{2015}]{colina15} Colina L., Piqueras-Lopez J., Arribas S., R. Riffel, Rodriguez-Ardila, Pastoriza, M.~G., Storchi-Bergmann T., Alonso-Herrero \& Sales D., 2015, A\&A, 578, 48.

\bibitem[\protect\citeauthoryear{Comer{\'o}n et al.}{2010}]{comeron} Comer{\'o}n, S., Knapen, J. H., Beckman, J. E., Laurikainen, E., Salo, H., Martínez - Valpuesta, I., Buta, R. J. 2010 MNRAS, 402, 2462

\bibitem[\protect\citeauthoryear{Cort{\'e}s, Kenney \& Hardy}{2015}]{cortes} Cort{\'e}s, J. R., Kenney, J. D. P., Hardy, E., 2015, ApJ\&SS, 216, 9

\bibitem[\protect\citeauthoryear{Couto et al.}{2013}]{couto2013} Couto, Guilherme S., Storchi-Bergmann T., Axon, David J., Robinson, A., Kharb, P., and Riffel, R. A., 2013, MNRAS, 435, 2982


\bibitem[\protect\citeauthoryear{de Vaucouleurs et al.}{1980}]{buta} de Vaucouleurs, G., Buta, R. J., AJ, 85, 637

\bibitem[\protect\citeauthoryear{de Vaucouleurs et al.}{1991}]{devaucouleurs91} de Vaucouleurs, G., de Vaucouleurs, A., Corwin, H. G. Jr, Buta, R. J., Paturel, G., Fouque, P., 1991, Third Reference Catalogue of Bright Galaxies, Vols 1--3, Springer-Verlag, Berlin

\bibitem[\protect\citeauthoryear{Dicaire et al.}{2008}]{dicaire} Dicaire, I., Carignan, C., Amram, P., Hernandez, O., Chemin, L., Daigle, O., de Denus-Baillargeon, M.-M., Balkowski, C., Boselli, A., Fathi, K., Kennicutt, R. C., 2008 MNRAS, 385, 553.

\bibitem[\protect\citeauthoryear{Dors et al.}{2008}]{dors08} Dors, O. L., Jr., Storchi-Bergmann, T., Riffel, R. A., Schimdt, Alex. A., 2008, A\&A, 428, 59.

\bibitem[\protect\citeauthoryear{Elmegreen \& Elmegreen}{1987}]{elmegreen87} Elmegreen D. M., Elmegreen B. G., 1987, ApJ, 314, 3

\bibitem[\protect\citeauthoryear{Elmegreen et al.}{1999}]{elmegreen99} Elmegreen D. M., Chromey F. R., Bissell B. A., Corrado K., 1999, AJ, 118,
2618

\bibitem[\protect\citeauthoryear{Emsellem et al.}{2003}]{emsellem} Emsellem, E., Goudfrooij, P., \& Ferruit, P. 2003, MNRAS, 345, 1297

\bibitem[\protect\citeauthoryear{Emsellem et al.}{2006}]{emsellem2006} Emsellem, E., Fathi, K., Wozniak, H., Ferruit, P., Mundell, C. G., Schinnerer, E. 2006, MNRAS, 365, 367

\bibitem[\protect\citeauthoryear{Emsellem et al.}{2015}]{emsellem15} Emsellem, E., Renaud, F., Bournaud, F., Elmegreen, B., Combes, F., Gabor, J. M., 2015, MNRAS, 446, 2468

\bibitem[\protect\citeauthoryear{Englmaier \& Shlosman}{2004}]{engl2004} Englmaier, P., \& Shlosman, I. 2004, ApJ, 617, L115

\bibitem[\protect\citeauthoryear{Elmegreen et al.}{2002}]{elmegreen} Elmegreen , D. M. and Elmegreen, B. G. and Eberwein, K. S., 2002 ApJ, 564, 243

\bibitem[\protect\citeauthoryear{Erwin \& Sparke}{1999}]{erwin} Erwin, P., Sparke L. S., 1999, ASP Conf. Ser. Vol. 182. Galaxy Dinamics - A 
Rutgers Symposium. Astron. Soc. Pac., San Francisco, p. 243

\bibitem[\protect\citeauthoryear{Erwin \& Sparke}{2002}]{erwin2002} Erwin, P., Sparke L. S., 2002, AJ, 124, 65

\bibitem[\protect\citeauthoryear{Erwin \& Sparke}{2003a}]{erwin03} Erwin, P., Sparke L. S., 2003, ApJS, 146, 299

\bibitem[\protect\citeauthoryear{Erwin \& Sparke}{2003b}]{erwin03b} Erwin, P., Beltr{\'a}n, J. C. V., Graham, A. W. and Beckman, J. E., 2003, ApJ, 597, 947

\bibitem[\protect\citeauthoryear{Fathi et al.}{2006}]{fathi06} Fathi, K., Storchi-Bergmann, T., Riffel, R. A., Winge, C., Axon, D. J., Robinson, A., Capetti, A., \& Marconi, A., 2006, ApJ, 641, L25.

\bibitem[\protect\citeauthoryear{Feltre, Chalrlot \& Gutkin}{2016}]{feltre16} Feltre, A., Charlot, S., Gutkin, J., 2016, 456, 3354. 

\bibitem[\protect\citeauthoryear{Gonzalez-Martin et al.}{2009}]{gonzalez-Martin09} Gonzalez-Martin, O., Masegosa, J., Marquez, I., Guainazzi, M., Jimenez-Bailon, E. 2009, A\&A, 506, 1107.

\bibitem[\protect\citeauthoryear{Helfer et al.}{2003}]{helfer03} Helfer, T. T., Thornley, M. D., Regan, M. W., Wong, T., Sheth, K., Vogel, S. N., Blitz, L., Bock, D. C.-J.,
 2003, ApJS, 145, 259.

\bibitem[\protect\citeauthoryear{Ho et al.}{1997}]{ho97} Ho, L. C., Filippenko, A. V., Sargent, W. L. 1997, ApJ\&SS, 112, 315.

\bibitem[\protect\citeauthoryear{Ho}{1999}]{ho99} Ho, L. C. 1999, ApJ, 516, 672.

\bibitem[\protect\citeauthoryear{Ho et al.}{2000}]{ho00} Ho, L., Rudnick, G., Rix, H-W, Shields, J. C., McIntosh, D. H., Filippenko, A. V., Sargent, W. L. W., Eracleous, M.,  2000, ApJ, 541, 1.

\bibitem[\protect\citeauthoryear{Ho \& Ulvestad}{2001}]{ho01} Ho, L. C., Ulvestad, J.S., 2001, ApJS, 133, 77.

\bibitem[\protect\citeauthoryear{Hook et al.}{2004}]{hook04} Hook, I., Jorgensen, I., Allington-Smith, J. R., Davies, R. L., Metcalfe, N., Murowinski, R. G., Crampton, D., 2004, PASP, 116, 425

\bibitem[\protect\citeauthoryear{Kauffmann et al.}{2003}]{kauffmann03} Kauffmann, G., Heckman, T. M., Tremonti, C., Brinchmann, J.,
Charlot, S., White, S. D. M., Ridgway, S. E., Brinkmann, J., Fukugita, M., Hall, P. B., Ivezi, Å., Richards, G. T., Schneider, D. P., 2003, MNRAS, 346, 1055

\bibitem[\protect\citeauthoryear{Kewley, Heisler \& Dopita }{2001}]{kewley01} Kewley, L. J., Heisler, C. A., Dopita, M. A., 2001, ApJS, 132, 37.

\bibitem[\protect\citeauthoryear{Kewley et al. }{2006}]{kewley06}  Kewley, L. J., Groves, B., Kauffmann, G., Heckman, T., 2006, MNRAS, 372, 961

\bibitem[\protect\citeauthoryear{Knapen et al.}{2002}]{knapen02} Knapen, J. H., Pérez-Ramírez, D., Laine, S., 2002, MNRAS 337, 808. 

\bibitem[\protect\citeauthoryear{Knapen}{2005}]{knapen} Knapen, J. H. 2005, ApJ\&SS 295, 85.

\bibitem[\protect\citeauthoryear{Koopmann, Kenney \& Young}{2001}]{koopmann01} Koopmann, R. A., Kenney, J. D. P., Young, J., 2001, ApJS 135, 125.

\bibitem[\protect\citeauthoryear{Kraemer et al.}{2011}]{kraemer} Kraemer, S. B., Schmitt, H. R., Crenshaw, D. M., Meléndez, M., Turner, T. J., Guainazzi M., Mushotzky, R. F., 2011, ApJ 727, 130 

\bibitem[\protect\citeauthoryear{Laine et al.}{2003}]{laine} Laine S., van der Marel R. P., Rossa J., Hibbard J. E., Mihos J. C., B\"oker T., Zabludoff A. I., 2003, AJ, 79, 745

\bibitem[\protect\citeauthoryear{Lena et al.}{2015}]{lena15} Lena, D. and Robinson, A.,  Storchi-Bergmann, T.,  Schnorr-M{\"u}ller, A.,  Seelig, T.,  Riffel, R.~A.,  Nagar, N.~M.,  Couto, G.~S. and Shadler, L.,  2015, ApJ, 806, 84


\bibitem[\protect\citeauthoryear{Lena et al.}{2016}]{lena16} Lena, D. and Robinson, A.,  Storchi-Bergmann, T., Couto, G.~S.,   Schnorr-M{\"u}ller, A.,  Riffel, R.~A.,  2016, MNRAS, 459, 4485.

\bibitem[\protect\citeauthoryear{Li et al.}{2011}]{li11} Li, J-T., Wang, Q. D., Li, Z., Chen, Y., 2011, ApJ, 737, 41.

\bibitem[\protect\citeauthoryear{Li, Shen \& Kim.}{2015}]{li15} Li, Z. Shen, J., Kim, W-T., 2015, ApJ, 806, 150. 
	

\bibitem[\protect\citeauthoryear{Liu}{2011}]{liu11} Liu, J., 2011, ApJS, 192, 10.

\bibitem[\protect\citeauthoryear{Maciejewski}{2002}]{maciejewski} Maciejewski, W., Teuben, P. J., Sparke, L. S., \& Stone, J. M. 2002, MNRAS, 329, 502.

\bibitem[\protect\citeauthoryear{Malkan, Gorjian \& Tam}{1998}]{malkan98} Malkan, M. A., Gorjian, V. \& Tam, R., 1998, ApJS, 117,25.

\bibitem[\protect\citeauthoryear{Markwardt et al.}{2009}]{mark09} Markwardt C. B., 2009, in Bohlender D. A., Durand D., Dowler P., eds, ASP
Conf. Ser. Vol. 411, Astronomical Data Analysis Software and Systems XVIII. Astron. Soc. Pac., San Francisco, p. 251

\bibitem[\protect\citeauthoryear{Martinsson et al.}{2013a}]{martinsson13} Martinsson, T. P. K., Verheijen, M. A. W., Westfall, K. B., Bershady, M. A., Schechtman - Root, A., Andersen, D. R., Swaters, R. A., 2013, A\&A, 557, A130

\bibitem[\protect\citeauthoryear{Martinsson et al.}{2013b}]{martinsson13b} Martinsson, T. P. K., Verheijen, M. A. W., Westfall, K. B., Bershady, M. A., Schechtman - Root, A., Andersen, D. R., Swaters, R. A., 2013, A\&A, 557, A131

\bibitem[\protect\citeauthoryear{Mazzalay et al.}{2013}]{mazzalay13} Mazzalay, X. et al., 2013, MNRAS, 428, 2389.

\bibitem[\protect\citeauthoryear{Mazzalay et al.}{2014}]{mazzalay14} Mazzalay, X. et al., 2014, MNRAS, 438, 2036.

\bibitem[\protect\citeauthoryear{Meléndez et al.}{2008}]{melendez} Meléndez, M., Kraemer, S. B., Schimitt, H. R., Crenshaw, D. M., Deo, R., P., Mushotzky, R. F., Bruhweiler, F. C., 2008, ApJ, 689, 95.

\bibitem[\protect\citeauthoryear{Munoz Marin et al.}{2007}]{munoz-marin07} Munoz Marin, V. M. et al., 2007, AJ, 134, 648.

\bibitem[\protect\citeauthoryear{Onodera}{2004}]{onodera} Onodera, S. and Koda, J. and Sofue, Y. and {Kohno}, K., 2004, PASJ, 56, 439

\bibitem[\protect\citeauthoryear{Osterbrock}{1989}]{oster89} Osterbrock D. E., 1989, Astrophysics of Gaseous Nebulae and Active Galactic Nuclei. University Science Books, Mill Valley, CA.

\bibitem[\protect\citeauthoryear{Paturel et al.}{2003}]{paturel03} Paturel, G., Petit, C., Prugniel, Ph., Theureau, G., Rousseau, J., Brouty, M., Dubois, P. \& Cambr\'esy, L., 2003, A\&A, 412, 45.

\bibitem[\protect\citeauthoryear{P\'erez-Ram\'irez et al.}{2000}]{perez-ramirez00}  P\'erez-Ram\'irez, D., Knapen, J. H., Peletier, R. F., Laine, S., Doyon, R., Nadeau, D., 2000, MNRAS, 317, 234.

\bibitem[\protect\citeauthoryear{Pogge \& Martini}{2002}]{pogge} Pogge R. W., Martini P., 2002, ApJ, 569, 624

\bibitem[\protect\citeauthoryear{Quillen et al.}{2001}]{quillen} Quillen, A. C., Alonso-Herrero, A., Lee, A., Shaked, S., Rieke, M. J., Rieke, G. H. 2001, ApJ, 547, 129

\bibitem[\protect\citeauthoryear{Regan, et al.}{1999}]{regan99} Regan, M. W. et al., 1999, ApJ, 117, 2676.


\bibitem[\protect\citeauthoryear{Rembold et al.}{2016}]{rembold16} Rembold, S. et al., 2016, MNRAS, submitted.

\bibitem[\protect\citeauthoryear{Repetto et al.}{2016}]{repetto} Repetto, P., Faúndez-Abans, M., Freitas-Lemes, P., Rodrigues, I., de Oliveira-Abans, M., 2016, MNRAS, 464, 293 

\bibitem[\protect\citeauthoryear{Reunanen, Kotilainen \& Prieto}{2002}]{reunanen02} Reunanen, J., Kotilainen, J. K., \& Prieto, M. A., 2002, MNRAS, 331, 154. 

\bibitem[\protect\citeauthoryear{Riffel et al.}{2006b}]{riffel2006b} Riffel, Rogemar A., Rodríguez-Ardila A., Pastoriza M. G., 2006b, A\&A, 457, 61.

\bibitem[\protect\citeauthoryear{Riffel et al.}{2008}]{riffel2008} Riffel, Rogemar A., Storchi-Bergmann, T., Winge, C., McGregor, P. J., Beck, T., Schmitt, H. 2008, MNRAS, 385, 1129.

\bibitem[\protect\citeauthoryear{Riffel et al.}{2010}]{profit} Riffel R. A., 2010, Ap\&SS, 327, 239.

\bibitem[\protect\citeauthoryear{Riffel, Storchi-Bergmann \& Nagar}{2010}]{mrk1066-exc} Riffel, Rogemar A., Storchi-Bergmann, T. \& Nagar, N. M., 2010a, MNRAS, 404, 166.

\bibitem[\protect\citeauthoryear{Riffel et al.}{2010}]{mrk1066_pop} Riffel, Rogemar A. \& Storchi-Bergmann, T., Riffel, R., \& Pastoriza, M. G., 2010, ApJ, 713, 469.

\bibitem[\protect\citeauthoryear{Riffel \& Storchi-Bergmann}{2011a}]{mrk1066c} Riffel, Rogemar A. \& Storchi-Bergmann, T., 2011, MNRAS, 411, 469.

\bibitem[\protect\citeauthoryear{Riffel, Storchi-Bergmann \& Winge}{2013}]{mrk79} Riffel, R. A., Storchi-Bergmann, T., Winge, C., 2013, 430, 2249.

\bibitem[\protect\citeauthoryear{Riffel et al.}{2015}]{n5929} Riffel, Rogemar A. \& Storchi-Bergmann, T. \& Riffel, R., 2015, MNRAS, 451, 3587.

\bibitem[\protect\citeauthoryear{Riffel et al.}{2017}]{llp_stel} Riffel, Rogemar A., Storchi-Bergmann, T., Riffel, R., Dahmer-Hahn, L. G., Diniz, M. R., Sch\"onell, A. J., Dametto, N. Z., 2017, MNRAS, submitted.



\bibitem[\protect\citeauthoryear{Riffel et al.}{2011}]{mrk1157_pop} Riffel, R., Riffel, Rogemar A., Ferrari, F., \& Storchi-Bergmann, T., 2011, MNRAS, 416, 493.


\bibitem[\protect\citeauthoryear{Rodr\'\i guez-Ardila et al.}{2004}]{ardila04} Rodr\'\i guez-Ardila, A.,  Pastoriza, M. G., Viegas, S., Sigut, T. A. A., \& Pradhan, A. K., 2004,  A\&A, 425, 457.

\bibitem[\protect\citeauthoryear{Sakamoto et al.}{1999a}]{sakamoto} Sakamoto K., Okumura S. K., Ishizuki S., Scoville N. Z., 1999, ApJ, 525, 691

\bibitem[\protect\citeauthoryear{Sanchez et al.}{2015}]{sanchez15} Sanchez, S. et. al. 2015, A\&A, 574, A47.

\bibitem[\protect\citeauthoryear{Sarzi et al.}{2010}]{sarzi10} Sarzi, M., et. al. 2010, MNRAS, 402, 2187.

\bibitem[\protect\citeauthoryear{Sarzi et al.}{2001}]{sarzi} Sarzi, M.,Sarzi, M., Rix, H.-W., Shields, J. C., Rudnick, G., Ho, L. C., McIntosh,
D. H., Filippenko, A. V., \& Sargent, W. L. W. 2001, ApJ, 550, 65

\bibitem[\protect\citeauthoryear{Schnorr--M{\"u}ller et al.}{2011}]{allan11} Schnorr M{\"u}ller A., Storchi-Bergmann T., Riffel R. A., Ferrari F., Steiner J. E., Axon D. J., Robinson A., 2011, MNRAS, 413, 149

\bibitem[\protect\citeauthoryear{Schnorr--M{\"u}ller et al.}{2014a}]{allan14} Schnorr M{\"u}ller A., Storchi-Bergmann T., Nagar, N. M., \& Ferrari, F. 2014a, MNRAS, 438, 3322


\bibitem[\protect\citeauthoryear{Schnorr--M{\"u}ller et al.}{2014b}]{allan14b} Schnorr M{\"u}ller A., Storchi-Bergmann T., Nagar, N. M., \& Robinson, A., Lena, D., Riffel, R. A., Couto, G. S., 2014b, MNRAS, 437, 1708.

\bibitem[\protect\citeauthoryear{Shlosman et al.}{1990}]{shlosman} Shlosman, I., Begelman, M. C., \& Frank, J. 1990, Nature, 345, 679

\bibitem[\protect\citeauthoryear{Shostak}{1987}]{shostak87} Shostak, G. S. 1987, A\&A, 175, 4.

\bibitem[\protect\citeauthoryear{Sim{\~o}es Lopes et al.}{2007}]{slopes} Sim{\~o}es Lopes, R.~D., Storchi-Bergmann, T., de F{\'a}tima Saraiva, M. and Martini, P., 2007, ApJ, 655, 718

\bibitem[\protect\citeauthoryear{Storchi-Bergmann et al.}{2007}]{thaisa07} Storchi-Bergmann, T., Dors Jr., O.,  Riffel,  R. A., Fathi, K.,  Axon, D. J., \& Robinson, A., 2007, ApJ, x,  x

\bibitem[\protect\citeauthoryear{Sutherland et al.}{1993}]{sutherland93} Sutherland, R. S., Bicknell, G. V., Dopita, M. A., 1993, ApJ,  414, 510.


\bibitem[\protect\citeauthoryear{Trippe et al.}{2010}]{trippe} Trippe, M. L., Crenshaw, D. M., Deo, R. P., Dietrich, M. Kraemer, S. B., Rafter, S. E., Turner, T. J., ApJ, 725, 1749

\bibitem[\protect\citeauthoryear{van der Kruit \& Allen}{1978}]{van} van der Kruit, P. C., \& Allen, R. J. 1978, ARA\&A, 16, 103

\bibitem[\protect\citeauthoryear{V{\'e}ron-Cetty \& Veron}{2006}]{veron} V{\'e}ron-Cetty, M. P., \& V{\'e}ron, P., 2006, A\&A, 455, 773.

\bibitem[\protect\citeauthoryear{Viegas \& Contini}{1994}]{viegas94} Viegas, S., Contini, M., 1994, ApJ, 428, 113.


\bibitem[\protect\citeauthoryear{Young et al.}{1996}]{young96}  Young, J. S., Allen, L., Kenney, J. D. P., Lesser, A., Rownd, B., 1996, AJ, 112, 1903.

\bibitem[\protect\citeauthoryear{Westfall et al.}{2011}]{westfall11} Westfall, K. B., Bershady, M. A., Verheijen, M. A. W., 2011, ApJS, 193, 21.	


\end{thebibliography}
\end{document}